\documentclass[10pt,aps,prd,superscriptaddress,nobibnotes,longbibliography,floatfix,twocolumn]{revtex4-2}

\usepackage{bm}
\usepackage{mathtools,
amsmath,
amssymb,
amsfonts,
mathrsfs,
chngcntr,
multirow}

\let\cc\corresponds
\let\corresponds\relax
\usepackage{mathabx}
\let\corresponds\cc

\usepackage[utf8]{inputenc}
\usepackage[T1]{fontenc}

\usepackage{soul}

\usepackage{amsmath}

\usepackage[dvipsnames]{xcolor}
\usepackage[unicode]{hyperref}
\hypersetup{colorlinks=true, citecolor=MidnightBlue,
            linkcolor=MidnightBlue, urlcolor=MidnightBlue, linktocpage=true}
\usepackage[normalem]{ulem}

\usepackage{graphicx}

\usepackage{tikz}
\usetikzlibrary{decorations.pathmorphing}

\newcommand{\D}{\mathrm{d}}

\begin{document}
\title{Singularity resolution and inflation\\from an infinite tower of regularized curvature corrections}

\author{Pedro G. S. Fernandes}
\email{fernandes@thphys.uni-heidelberg.de}
\affiliation{Institut f\"ur Theoretische Physik, Universit\"at Heidelberg, Philosophenweg 12, 69120 Heidelberg, Germany}

\begin{abstract}
    We explore four-dimensional scalar-tensor theories obtained from well-defined dimensional regularizations of Lovelock invariants. When an infinite tower of corrections is considered, these theories allow for cosmological models in which the Big Bang singularity is replaced by an inflationary phase in the early-universe, and they also admit a specific class of regular black hole solutions.
\end{abstract}

\maketitle

Einstein's theory of General Relativity (GR) breaks down at very high energies. Under standard assumptions, it predicts the existence of singularities \cite{Penrose:1964wq,Hawking:1967ju,Hawking:1970zqf}. This occurs, for instance, deep in the cores of black holes and in the very beginning of the universe.

The Einstein-Hilbert action is widely expected to be only the leading term in an infinite series of higher-curvature corrections that become important at sufficiently high energy scales. This expectation naturally arises from the Wilsonian low-energy approach and is realized, for instance, in leading candidates for quantum theories of gravity, such as string theory \cite{Gross:1986mw,Gross:1986iv,Grisaru:1986vi} and asymptotically safe quantum gravity \cite{Bonanno:2020bil,Knorr:2022dsx,Knorr:2018kog,Dupuis:2020fhh}.

Recent work has shown that regular black holes solutions exist \cite{Bueno:2024dgm} and can form dynamically \cite{Bueno:2024zsx,Bueno:2024eig} in higher dimensions, $D \geq 5$, within the framework of quasi-topological gravity \cite{aguilar-gutierrezAspectsHighercurvatureGravities2023, ahmedQuintessentialQuarticQuasitopological2017, buenoFourdimensionalBlackHoles2017, buenoGeneralizedQuasitopologicalGravities2019, buenoGeneralizedQuasitopologicalGravities2022, buenoUniversalBlackHole2017, hennigarBlackHolesEinsteinian2016, hennigarGeneralizedQuasitopologicalGravity2017, morenoClassificationGeneralizedQuasitopological2023a,Moreno:2023arp,Cisterna:2018tgx} when accounting for an infinite tower of curvature corrections. This represents a major step forward in tackling one of the key challenges in theoretical physics: identifying a mechanism to resolve spacetime singularities \cite{Carballo-Rubio:2025fnc}. Unlike many phenomenological regular black hole models, this approach avoids the need for \textit{ad hoc} exotic matter or theories that impose fine-tuned relationships between a solution’s integration constants and the theory’s parameters, as often seen in non-linear electrodynamics models \cite{Ayon-Beato:1998hmi,Ayon-Beato:2000mjt,Bronnikov:2022ofk}.
In cosmological settings, it has been shown that, in a similar framework, infinite towers of curvature corrections \cite{Moreno:2023arp} can replace the initial singularity with an inflationary period \cite{Arciniega:2018tnn}. This framework is known as ``geometric inflation'' \cite{Arciniega:2018tnn,Arciniega:2018fxj,Arciniega:2020pcy,Edelstein:2020nhg,Jaime:2021pqn,Arciniega:2025viq}, see also e.g. Refs. \cite{Bojowald:2002nz, deCesare:2016axk, Starobinsky:1980te} for other frameworks where inflation is achieved purely from gravitational effects. Unfortunately, these theories generally lead to field equations of higher-order in derivatives, except in specific backgrounds, making them susceptible to Ostrogradsky instabilities \cite{BeltranJimenez:2023mxp,DeFelice:2023vmj}, and quasi-topological gravities do not exist in four-dimensions \cite{buenoGeneralizedQuasitopologicalGravities2022}.

Lovelock theories of gravity \cite{Lovelock:1971yv,Padmanabhan:2013xyr} generalize GR to higher dimensions as the unique class of purely metric, local, and diffeomorphism-invariant theories that maintain second-order equations of motion. Therefore, despite the action containing higher-curvature invariants, these theories avoid Ostrogradsky instabilities. However, in a spacetime with $D$ dimensions, only up to $\lceil D/2 \rceil$ non-trivial Lovelock invariants can be included in the gravitational action, with all others being either topological or vanishing. Consequently, within this framework, four-dimensional theories cannot accommodate an infinite tower of Lovelock curvature corrections, as the action is restricted to the Einstein-Hilbert term plus a cosmological constant.

In recent years, there has been a surge of interest in the quadratic Lovelock invariant -- the Gauss-Bonnet term -- and the possibility of a non-trivial Gauss-Bonnet–corrected theory of gravity in four-dimensions. This idea was first introduced in Ref. \cite{Glavan:2019inb} through a singular rescaling of the Gauss-Bonnet coupling constant in a four-dimensional limit of the higher-dimensional theory. However, this singular limit was later shown to be ill-defined \cite{Gurses:2020ofy}, leading to the development of well-defined regularizations. These regularizations resulted in specific four-dimensional Horndeski scalar-tensor theories \cite{Fernandes:2020nbq, Hennigar:2020lsl, Lu:2020iav, Kobayashi:2020wqy, Fernandes:2021dsb}, which preserve many solutions and properties of the higher-dimensional Einstein-Gauss-Bonnet theory. As a result, they have been extensively studied; see Ref. \cite{Fernandes:2022zrq} for a comprehensive review.

In this work, we take a more general approach by considering Lovelock-like corrections to all orders in curvature as well-defined four-dimensional scalar-tensor theories and investigating their implications in both cosmology and black hole physics for the first time. In the limit where an infinite tower of corrections is taken into account, we show that the Big Bang singularity is replaced by a period of inflation in the early universe, and that a specific class of regular black hole solutions exist. Throughout, we adopt units where $c = G = 1$.

\noindent \textbf{\textit{Scalar-Tensor Theories from Regularized Lovelock Gravity.}}
The $n^{\rm th}$ order Lovelock invariant is given by
\begin{equation}
    \mathcal{R}^{(n)} \equiv \frac{1}{2^n} \delta^{\mu_1 \nu_1 \dots \mu_n \nu_n}_{\alpha_1 \beta_1 \dots \alpha_n \beta_n} \prod_{i=1}^{n} R^{\alpha_i \beta_i}_{\phantom{\alpha_i \beta_i} \mu_i \nu_i},
\end{equation}
where $\delta^{\mu_1 \nu_1 \dots \mu_n \nu_n}_{\alpha_1 \beta_1 \dots \alpha_n \beta_n} \equiv n! \delta^{\mu_1}_{[\alpha_1} \delta^{\nu_1}_{\beta_1} \dots \delta^{\mu_n}_{\alpha_n} \delta^{\nu_n}_{\beta_n]}$ is the generalized Kronecker delta, and $n\geq 0$. The $D$-dimensional Lovelock Lagrangian is composed by a linear combination of the first $\lceil D/2 \rceil$ Lovelock invariants, as the higher-order invariants either become topological or vanish. In particular, in four-dimensions we recover GR with a cosmological constant.

To obtain Lovelock-like corrections at all orders curvature in four-dimensions, we apply a well-defined dimensional regularization to each Lovelock invariant, resulting in scalar-tensor theories within the Horndeski class \cite{Horndeski:1974wa, Kobayashi:2019hrl}. This regularization method was first introduced in Ref. \cite{Mann:1992ar} to recover GR-like dynamics in two dimensions by regularizing the Ricci scalar and was later extended in Refs. \cite{Fernandes:2020nbq, Hennigar:2020lsl} to formulate a well-defined Gauss-Bonnet theory in four-dimensions. This method relies on the use of two conformally related metrics, $\tilde{g}_{\mu \nu} = e^{-2\phi} g_{\mu \nu}$, and the following limit at each Lovelock order
\begin{equation}
    \sqrt{-g} \mathcal{L}^{(n)} = \lim_{d\to 2n} \frac{\sqrt{-g}\mathcal{R}^{(n)} - \sqrt{-\tilde{g}}\tilde{\mathcal{R}}^{(n)}}{d-2n},
    \label{eq:limit}
\end{equation}
where the expressions inside the limit are to be evaluated in $d$-dimensions, and the limit is taken as $d$ approaches the \textit{critical} dimension $2n$, where the $n^{\rm th}$ Lovelock invariant becomes topological. This computation for $n=1$ in $2D$, and $n=2$ in $4D$ was performed in detail in Ref. \cite{Fernandes:2020nbq}, and it can be shown that the limit is well-defined for every $n$, resulting in a scalar-tensor theory, with scalar $\phi$, and second-order equations of motion \cite{Colleaux:2020wfv}. Although the limit is taken as $d \to 2n$, the resulting scalar-tensor Lagrangian can be evaluated within a four-dimensional action for any $n$. This four-dimensional theory is inspired by the three-dimensional limits of Gauss-Bonnet gravity and their solutions \cite{Hennigar:2020fkv, Hennigar:2020drx}, which, despite not being considered in the Gauss-Bonnet critical dimension $D = 4$, lead to solutions similar to those in higher-dimensional Gauss-Bonnet gravity. The final result at each $n$ for the limit \eqref{eq:limit} is that $\mathcal{L}^{(n)}$ is proportional to a four-dimensional Horndeski Lagrangian
\begin{equation}
    \begin{aligned}
		&\mathcal{L}_H = G_2(\phi,X)-G_3(\phi,X)\Box\phi + G_4(\phi,X)R
		\\&
        +G_{4X}  \left[(\Box\phi)^2-\left(\nabla_\mu \nabla_\nu \phi\right)^2\right]
          +G_5(\phi,X) G^{\mu\nu}\nabla_\mu \nabla_\nu \phi
		  \\&
          -\frac{G_{5X}}{6}\left[
          (\Box\phi)^3-3\Box\phi \left(\nabla_\mu \nabla_\nu \phi\right)^2
          +2\left(\nabla_\mu \nabla_\nu \phi\right)^3
          \right],
    \end{aligned}
	\label{eq:Horndeski}
\end{equation}
with functions given by \cite{Colleaux:2020wfv}
\begin{equation}
    \begin{aligned}
        &G_2^{(n)} = 2^{n+1}(n-1)(2n-3) X^n,\\&
        G_3^{(n)} = -2^{n} n(2n-3) X^{n-1},\quad
        G_4^{(n)} = 2^{n-1} n X^{n-1},\\&
        G_5^{(n)} = -\begin{cases} 4 \log X, \qquad n=2, \\ 2^{n-1} \frac{n (n-1)}{n-2} X^{n-2}, \qquad n > 2, \end{cases}
        \label{eq:HorndeskiFunctions}
    \end{aligned}
\end{equation}
where $X = -\partial_\mu \phi \partial^\mu \phi/2$.
The gravitational action studied in this work is  
\begin{equation}  
    S = \frac{1}{16\pi} \int \mathrm{d}^4 x \sqrt{-g} \left[ -2 \Lambda + R + \frac{1}{\ell^2}\sum_{n=2}^{\infty} c_n \ell^{2n} \mathcal{L}^{(n)} \right],  
    \label{eq:action}  
\end{equation}  
where $\{c_n\}$ are dimensionless coupling constants, $\ell$ is a new length scale, and now $\mathcal{L}^{(n)}$ is the Horndeski Lagrangian with functions given by Eq. \eqref{eq:HorndeskiFunctions}. This theory falls within the shift-symmetric class of Horndeski theories, see e.g. \cite{Pujolas:2011he,Deffayet:2010qz,Germani:2017pwt,BorislavovVasilev:2024loq,Creminelli:2020lxn,Sotiriou:2013qea,Sotiriou:2014pfa,Delgado:2020rev,Khoury:2020aya,Saravani:2019xwx,Babichev:2017guv}, implying the existence of a conserved current $J^\mu$, whose expression is provided in Ref. \cite{Saravani:2019xwx}. The vanishing divergence of this current, $\nabla_\mu J^\mu = 0$, is equivalent to the scalar field equation of motion. Importantly, a sufficient condition to solve the scalar field equation of motion, is $J^\mu = 0$. The field equations for a generic Horndeski Lagrangian \eqref{eq:Horndeski} are presented in Refs.~\cite{Kobayashi:2011nu,Lecoeur:2024kwe}.

When the whole tower of corrections is considered, the theory described by Eq. \eqref{eq:action} can in some cases be re-summed into Horndeski theories with a non-local structure. Examples are presented in the Supplemental Material.

\begin{figure*}[]
	\centering
	\includegraphics[width=0.5\linewidth]{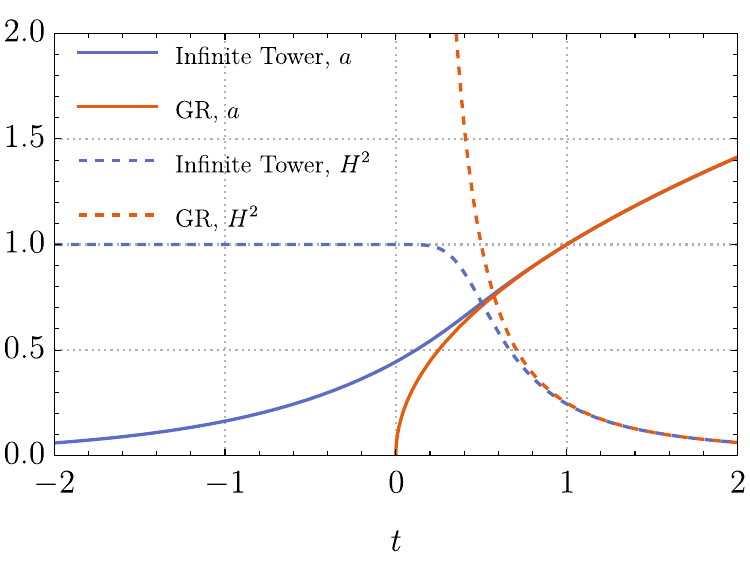}\hfill
	\includegraphics[width=0.5\linewidth]{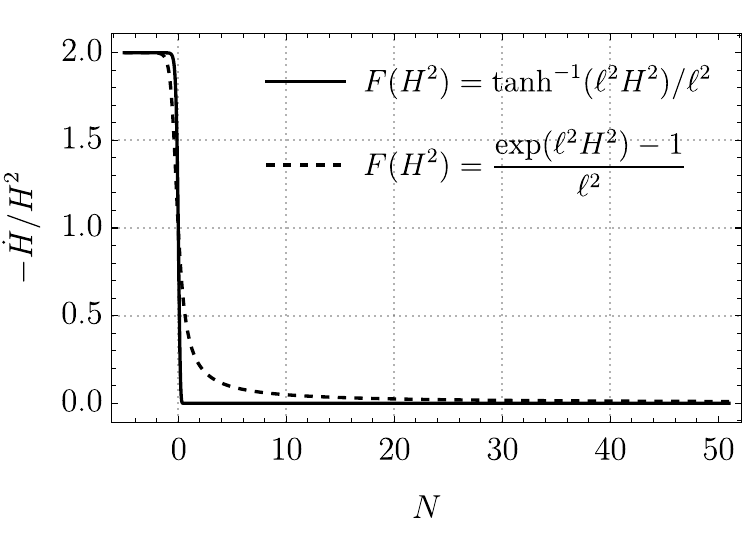}\vfill
    \caption{(Left) Time evolution of $a$ and $H^2$ (in units $\ell=1$) for the modified Friedmann equation in Eq. \eqref{eq:FriedmannExample}, and in GR, assuming a radiation dominated universe with initial value for the energy density $\rho = 3/(32\pi)$ and scale factor $a=1$ at time $t=1$. In GR, a Big Bang singularity occurs at $t=0$, whereas when an infinite tower of corrections is considered, the singularity is replaced by an inflationary phase. (Right) Slow-roll parameter $-\dot H/H^2$ as a function of the number of e-folds $N$ (defined such that $N=0$ at the end of inflation, i.e., evolution proceeds from right to left), for the theories with couplings given by $c_n = (1-(-1)^n)/(2n)$, and $c_n=1/n!$. The curves are shown up to $N=50$, but they continue all the way to $N\to \infty$.}
	\label{fig:hubble}
\end{figure*}

\noindent \textbf{\textit{Inflation replaces the Big Bang singularity.}}
Consider a homogeneous, isotropic, and spatially-flat Friedmann-Lemaître-Robertson-Walker (FLRW) line-element, given by
\begin{equation}  
    \D s^2 = - \D t^2 + a(t)^2 \left( \D r^2 + r^2 \D \theta^2 + r^2 \sin^2 \theta \D \varphi^2 \right),
    \label{eq:flrw}  
\end{equation}  
where $a(t)$ is the scale factor. The scalar field is assumed to be only a function of $t$. Time derivatives are denoted by an overdot, and we introduce the Hubble parameter $H = \dot a/a$. We consider a perfect fluid stress-energy tensor for matter, $T^{\mu}_{\phantom{\mu}\nu} = \mathrm{diag}\left( -\rho, p, p, p \right)$, where the energy density $\rho$ and pressure $p$ obey the continuity equation $\dot \rho + 3 H \left( \rho + p \right) = 0$.

For the theory \eqref{eq:action} on an FLRW background, the only non-trivial component of the shift-symmetry current is $J^t$. The scalar field equation can be integrated to the form $J^t = Q/a^3$, where $Q$ is an integration constant interpreted as the shift-symmetry scalar charge, and $J^t$ is interpreted as the scalar charge density.
At each order $n$, the current vanishes if the scalar field obeys
\begin{equation}
    \phi = \log \left(a/a_0\right),
    \label{eq:scalarflrw}
\end{equation}
where $a_0$ is an arbitrary integration constant. This profile solves the scalar field equation of motion for any $n$, implying $Q = 0$. While we were unable to find exact solutions for $Q \neq 0$, solutions with a vanishing charge/current are well-motivated for several reasons: (i) they are attractors in an expanding universe; (ii) the scalar adopts the profile \eqref{eq:scalarflrw}, which aligns with expectations from the higher-dimensional origin of the theory, where it represents the logarithm of the scale factor of the extra-dimensional space prior to taking the 4D limit \cite{Lu:2020iav,Kobayashi:2020wqy}; (iii) the absence of additional charges is expected when interpreting the theory as a 4D Lovelock theory; and (iv) as shown below, it leads to Friedmann equations that match those of Lovelock gravity \cite{Deruelle:1989fj}.

Using the scalar field profile \eqref{eq:scalarflrw} in the Einstein equations, we obtain the generalized Friedmann equation
\begin{equation}
    F(H^2) = \frac{8\pi}{3}\rho + \frac{\Lambda}{3}.
    \label{eq:friedmann}
\end{equation}
where we have defined the function\footnote{$c_1=1$ is the coefficient of the Einstein-Hilbert term.} 
\begin{equation}
    F(x) = \frac{1}{\ell^2}\sum_{n=1}^{\infty} c_n (\ell^2 x)^{n}.
    \label{eq:function}
\end{equation}
As in Lovelock gravity \cite{Deruelle:1989fj}, the $n^{\text{th}}$ order contribution to the Friedmann equation is proportional to $H^{2n}$. As an example, for the theory with $c_n = \left(1-(-1)^n\right)/(2n)$, we obtain $F(H^2) = \tanh^{-1}\left(\ell^2 H^2\right)/\ell^2$
such that the Friedmann equation can be written as
\begin{equation}
    H^2 = \frac{1}{\ell^2} \tanh\left[ \ell^2\left(\frac{8\pi}{3}\rho + \frac{\Lambda}{3}\right) \right].
    \label{eq:FriedmannExample}
\end{equation}
Examples for other choices of couplings are presented in the Supplemental Material.
In this scenario, as $\rho \to \infty $ in the very early universe\footnote{From the continuity equation, we obtain $\rho = \rho_0 a^{-3(1+\omega)}$, where $\omega$ is the equation of state parameter, $p = \omega \rho$. In the early universe, radiation ($\omega=1/3$) dominates over matter ($\omega=0$).}, the spacetime must asymptotically approach a non-singular de Sitter geometry with $H^2 \to 1/\ell^2$. Consequently, the Big Bang singularity is replaced by an inflationary phase in the early universe. This non-singular scenario differs from others, such as bounces, as there is no contracting phase, and the energy density is not bounded from above. However, infinite energy density is never attained in finite time. This inflationary phase can be observed in Fig. \ref{fig:hubble} (left) we compare the time evolution of $H^2$ and $a$ for a radiation dominated universe\footnote{The same qualitative conclusions hold when other initial conditions, and equations of state, such as matter ($\omega=0$), are considered.} in GR, and in the theory with Friedmann equation given in Eq. \eqref{eq:FriedmannExample}. For late times, the evolution of the universe is well-described by GR. Other couplings we have explored, such as $c_n=1$, $c_n = 1/n$, $c_n=1/n!$, and $c_n = 1/(n-1)!$, also replace the initial singularity with an inflationary phase. In Fig. \ref{fig:hubble} (right), we plot the slow-roll parameter $-\dot H/H^2$ as a function of the number of e-folds $N$, for the theories with $F(H^2) = \tanh^{-1}\left(\ell^2 H^2\right)/\ell^2$ and $F(H^2) = (\exp\left(\ell^2 H^2\right)-1)/\ell^2$. We observe an inflationary period ($-\dot H/H^2<1$) followed by a graceful transition into a radiation dominated universe ($-\dot H/H^2=2$).
As shown in Refs. \cite{Arciniega:2018tnn,Cisterna:2024ksz}, the singularity is avoided provided the infinite tower is included, with the scale factor exhibiting at least exponential growth in the far past. Thus, both singularity resolution and inflation emerge as generic features, largely insensitive to the specific couplings $c_n$ and the choice of initial conditions.

\noindent \textbf{\textit{Regular black holes.}}
Black holes are predicted to form through gravitational collapse. The simplest case, initially explored by Oppenheimer and Snyder \cite{PhysRev.56.455}, involves the collapse of a homogeneous and isotropic sphere of pressureless dust. The interior metric describing this collapsing dust, which must be matched to an exterior metric, can be represented by the FLRW line-element and its dynamics governed by the Friedmann equations. In GR, the dust collapses in finite time to form a singularity and a Schwarzschild black hole.
However, when the full tower of corrections in Eq. \eqref{eq:action} is taken into account, the Friedmann equations are modified. For the examples studied in the previous section, such as Eq. \eqref{eq:FriedmannExample}, the collapsing dust, in the marginally-bound case \cite{Cipriani:2024nhx}, never forms a singularity in finite time. Beyond the marginally-bound case, the interior geometry is described by a spatially curved FLRW metric. In the Supplemental Material, we present the modified Friedmann equations of the theory \eqref{eq:action} in this setting. Even in this case, $H$ remains bounded for the cases we studied, and the conclusions drawn from the spatially-flat case apply.
This raises questions about whether singularities can form at all through gravitational collapse in this framework.

Despite the uncertainty surrounding the formation of singularities through gravitational collapse in this class of theories, it remains a compelling question whether singularities are an inevitable feature of black hole solutions in such theories, as they are in GR.
For this purpose, we consider a line-element of the form
\begin{equation}
    \D s^2 = - f(t,r) N(t,r)^2 \D t^2 + \frac{\D r^2}{f(t,r)} + r^2 \D \Omega^2_k,
    \label{eq:SSS}
\end{equation}
where $k$ can take values $\{+1,0,-1\}$ corresponding to positive, zero and negative horizon curvature, respectively.
Similarly to Ref. \cite{Alkac:2022fuc}, which studied the cubic Lovelock case ($n=3$), and as detailed in the Supplemental Material, we have only been able to integrate the field equations in the case $k=0$, which we consider from now on. Planar black holes~\cite{Birmingham:1998nr,Lemos:1997bd,Lemos:1995cm}\footnote{For $k=0$, it is possible to have toroidal, cylindrical or planar topology, depending on the range of the coordinates defining the two-dimensional space $\D \Omega^2_0$ \cite{Lemos:1997bd}.} require a negative cosmological constant, $\Lambda=-3/L^2$, which we define in terms of the anti-de Sitter (AdS) radius $L$. The planar black hole solution in GR is given by \cite{Birmingham:1998nr} $N(t,r)=1$ and
\begin{equation}
    f(t,r)=f_{\rm GR}(r) = \frac{r^2}{L^2} - \frac{2M}{r},
    \label{eq:fGR}
\end{equation}
where $M$ is an integration constant related to the mass of the black hole. This metric has a curvature singularity at $r=0$, as seen by computing the Kretschmann scalar, $R_{\mu \alpha \nu \beta} R^{\mu \alpha \nu \beta} = 48 M^2/r^6 + 24/L^2$.

Considering now the theory defined in Eq. \eqref{eq:action}, we proceed analogously to the FLRW case. The shift-symmetry current, at each order $ n $, vanishes identically if  
\begin{equation}  
    \phi = \log (r/r_0),
    \label{eq:scalarSSS}
\end{equation}  
where $ r_0 $ is an arbitrary constant. For this scalar field profile, the scalar field equation is automatically satisfied. Substituting this scalar field profile in the Einstein equations, the $tr$ component imposes that $f(t,r)\equiv f(r)$ at all orders $n$. Then, a combination of the $tt$ and $rr$ field equations, imposes that $N(t,r) \equiv N(t)$, which can be absorbed into a redefinition of the time coordinate. Therefore, $N(t,r)=1$ without loss of generality. The system reduces to a single equation, that determines the metric function $f(r)$. At each order $ n $, this equation can be integrated into an algebraic equation 
\begin{equation}
    F\left(- \frac{f(r)}{r^2}\right) = -\frac{f_{\rm GR}(r)}{r^2},
    \label{eq:planarBHs}  
\end{equation}
where the function $F(x)$ was defined in Eq. \eqref{eq:function} and $f_{\rm GR}(r)$ in Eq. \eqref{eq:fGR}.
This leads us to the conclusion that the class of theories examined in this letter satisfy a staticity theorem\footnote{To our knowledge, this is the first example of a staticity theorem holding for a class of higher-derivative theories beyond GR in four-dimensions. However, see Refs.~\cite{Bueno:2025jgc,Oliva:2010eb,Oliva:2011xu,Bueno:2024eig,Bueno:2024zsx} for specific cases in $2+1$ dimensions and higher-dimensional scenarios where staticity (and in some cases, Birkhoff) theorems have been established.}: assuming the scalar field profile \eqref{eq:scalarSSS}, any solution of the form \eqref{eq:SSS} within the theory \eqref{eq:action} must be both static and unique, characterized by $N(t,r) = 1$ and determined by the solution to Eq. \eqref{eq:planarBHs}. This result can be used to study the collapse of matter and the dynamical formation of black holes, following e.g. the approach of Refs.~\cite{Bueno:2024eig,Bueno:2024zsx}. Note that the vanishing of $J^\mu$ is necessary, as the scalar field equation takes the form $J^r = Q/r^2$, and if $Q\neq 0$, the scalar quantity $J^\mu J_\mu = Q^2/(r^4 f(r))$ diverges at the horizon, and at $r=0$.

As discussed in Refs. \cite{dePaulaNetto:2023cjw, Lobo:2020ffi, Bolokhov:2024sdy, Simpson:2023apa}, for static spacetimes of the form \eqref{eq:SSS}, a necessary and sufficient condition for the regularity of all independent components of the Riemann tensor, $R^{\mu \nu}_{\phantom{\mu} \phantom{\nu} \alpha \beta}$, and consequently all curvature invariants constructed from it, is that the Kretschmann scalar remains finite everywhere. By analyzing the Kretschmann scalar, one can show that the singularity at $r = 0$ is resolved if the metric function behaves as $f(r) = -r^2/\lambda^2 + \mathcal{O}(r^3)$ near the origin, where $\lambda$ is a constant with dimensions of length.
From Eq. \eqref{eq:planarBHs}, we find that resolving the singularity is possible if the function $F(x)$ develops a pole as $x \to 1/\lambda^2$. Sufficient conditions for the existence of a pole are: $c_n \geq 0$ for all $n$; a non-zero finite radius of convergence $\mathfrak{R}=1/\lim_{n \to \infty} (c_n)^{1/n}$; the series $\sum_n c_n \mathfrak{R}^n$ diverges. This class of couplings resolve the singularity of both cosmological models and black holes. Since there are infinitely many possible choices of the coefficients $\{c_n\}$ that could induce such a pole in $F(x)$, we adopt as a representative example the same couplings used in the previous section, $c_n = \left(1 - (-1)^n\right) / (2n)$, which leads to a pole at $x \to 1/\ell^2$.
In this case, the solution to Eq. \eqref{eq:planarBHs} is
\begin{equation}  
    f(r) = \frac{r^2}{\ell^2} \tanh \left[ \frac{\ell^2}{r^2} f_{\rm GR}(r) \right],
    \label{eq:BHsol1}  
\end{equation}
where the GR solution is recovered in the limit $ \ell \to 0 $. Additional examples are provided in the Supplemental Material. The location of the event horizon remains unchanged from the GR case, $r_H=(2L^2 M)^{1/3}$, and the solution is asymptotically AdS. This metric function is everywhere smooth and non-divergent, even when analytically extended to negative values of the coordinate $r$ \cite{Zhou:2022yio,Zhou:2023lwc}. The solution is also free of standard curvature singularities as the Kretschmann scalar is everywhere bounded, obeying $\lim_{r\to 0} R_{\mu \alpha \nu \beta} R^{\mu \alpha \nu \beta} = 24/\ell^4$. Additional solutions which are non-singular at $r = 0$ can be found, for instance, in the theories where $c_n = 1$ or $c_n = 1/n$.

The metric function in Eq. \eqref{eq:BHsol1} satisfies $f(0) = 0$ and $f'(0) = 0$. Thus, $r = 0$ is a Killing horizon associated to the Killing vector $\xi^\mu  = (\partial_t)^\mu$, since $g_{\mu \nu} \xi^\mu \xi^\nu|_{r=0} = -f(0) = 0$. Moreover this horizon is extremal in the sense that it has zero surface gravity, $\kappa|_{r=0} = - f'(0)/2 = 0$. Details on the causal structure of the spacetime \eqref{eq:BHsol1} are given in the Supplemental Material. The vanishing of the surface gravity prevents classical mass inflation instabilities that plague most regular black hole models \cite{Carballo-Rubio:2022kad,Carballo-Rubio:2024dca,Franzin:2022wai}. Thus, an infinite tower of corrections not only cures the singularity, but also has the potential to prevent classical mass inflation instabilities. This situation is similar to that observed in some regular black holes in $2+1$ dimensions \cite{Bueno:2021krl,Bueno:2025jgc}.

The infinite tower of corrections is essential: truncating the series in Eq. \eqref{eq:planarBHs} at any finite order introduces a singularity in the solutions. This can be observed starting from Eq. \eqref{eq:planarBHs}, truncated at $n=n_{\rm max}$, from which we get that near $r=0$ the metric function $f(r)$ behaves as
\begin{equation}
    f(r) \approx - \left(\frac{2M}{c_{n_{\rm max}}}\right)^{1/n_{\rm max}} \frac{r^{2-3/n_{\rm max}}}{\ell^{2-2/n_{\rm max}}}.
\end{equation}
Regular solutions with $f = -r^2/\ell^2 + \mathcal{O}(r^3)$, can only be achieved in the limit $n_{\rm max} \to \infty$, provided that $\mathfrak{R}$ is finite.

\noindent \textbf{\textit{Discussion.}}
In this work, we have explored one of the most fundamental open problems in theoretical physics -- the singularity problem. By including an infinite tower of Lovelock-like corrections to GR through well-defined scalar-tensor Horndeski theories, we have derived solutions that replace the initial singularity with an inflationary phase in the early universe, followed by a graceful exit into standard GR evolution, and solutions that describe regular planar black holes free of singularities. These theories introduce a single additional length scale, $\ell$, which sets the scale for new physics, and crucially, they do not suffer from Ostrogradsky instabilities.

The results of this work qualitatively align with those of Ref.~\cite{Bueno:2024dgm}, which studied higher-dimensional quasi-topological gravity and showed that an infinite series of corrections to GR can act as a mechanism for resolving singularities. Despite being explored in different settings, both Ref. \cite{Bueno:2024dgm} and this work support the paradigm that infinite towers of corrections may be key to the absence of singularities. Such corrections are a common feature of leading candidates for a theory of quantum gravity. However, our findings suggest that knowledge of the full theory is essential, as truncating the series of corrections typically results in singular spacetimes.

There are several promising directions for future research. A crucial next step would be to generalize our results to black holes with spherical horizons, as these are the most astrophysically relevant solutions. While we have not yet found a way to integrate the field equations in this case, an extension of the theories considered here might make this possible -- see e.g. the class of theories in Refs. \cite{Fernandes:2021dsb,Lu:2020iav}, where a slightly different regularization procedure of the quadratic Lovelock invariant is considered. Exploring gravitational collapse scenarios, such as those of Refs. \cite{Bueno:2024zsx,Bueno:2024eig} -- see also Refs. \cite{Husain:2021ojz,Lewandowski:2022zce,Bonanno:2023rzk} -- would be an important avenue for further investigation.
Additionally, investigating the inflationary predictions of these theories, and their cosmological stability -- with respect to both spatial anisotropies \cite{deCesare:2020swb}, and to perturbations \cite{Kobayashi:2016xpl,Libanov:2016kfc} -- would be an interesting direction. The generic no-go theorems of Horndeski gravity \cite{Kobayashi:2016xpl,Libanov:2016kfc} are expected to apply in our case. Nevertheless, our construction can be extended beyond this framework to circumvent these theorems. For instance, introducing a distinct scalar field at each order in the Lovelock expansion yields the same phenomenology for both cosmology and black holes as the single scalar case studied in this work. It would also be interesting to study the black holes presented in this work from the point of view of thermodynamics and holography \cite{Hennigar:2017umz,Cadoni:2016hhd,Peca:1998dv,Hartnoll:2009sz,Hossenfelder:2014gwa}.

\noindent \textbf{\textit{Acknowledgments.}} P.F. thanks Vitor Cardoso, Christos Charmousis, Aimeric Colléaux, Astrid Eichhorn, and Mokhtar Hassaine for valuable discussions and comments on a version of the manuscript.
This work is funded by the Deutsche Forschungsgemeinschaft (DFG, German Research Foundation) under Germany’s Excellence Strategy EXC 2181/1 - 390900948 (the Heidelberg STRUCTURES Excellence Cluster).

\bibliography{biblio.bib}

\begin{thebibliography}{105}%
\makeatletter
\providecommand \@ifxundefined [1]{%
 \@ifx{#1\undefined}
}%
\providecommand \@ifnum [1]{%
 \ifnum #1\expandafter \@firstoftwo
 \else \expandafter \@secondoftwo
 \fi
}%
\providecommand \@ifx [1]{%
 \ifx #1\expandafter \@firstoftwo
 \else \expandafter \@secondoftwo
 \fi
}%
\providecommand \natexlab [1]{#1}%
\providecommand \enquote  [1]{``#1''}%
\providecommand \bibnamefont  [1]{#1}%
\providecommand \bibfnamefont [1]{#1}%
\providecommand \citenamefont [1]{#1}%
\providecommand \href@noop [0]{\@secondoftwo}%
\providecommand \href [0]{\begingroup \@sanitize@url \@href}%
\providecommand \@href[1]{\@@startlink{#1}\@@href}%
\providecommand \@@href[1]{\endgroup#1\@@endlink}%
\providecommand \@sanitize@url [0]{\catcode `\\12\catcode `\$12\catcode `\&12\catcode `\#12\catcode `\^12\catcode `\_12\catcode `\%12\relax}%
\providecommand \@@startlink[1]{}%
\providecommand \@@endlink[0]{}%
\providecommand \url  [0]{\begingroup\@sanitize@url \@url }%
\providecommand \@url [1]{\endgroup\@href {#1}{\urlprefix }}%
\providecommand \urlprefix  [0]{URL }%
\providecommand \Eprint [0]{\href }%
\providecommand \doibase [0]{https://doi.org/}%
\providecommand \selectlanguage [0]{\@gobble}%
\providecommand \bibinfo  [0]{\@secondoftwo}%
\providecommand \bibfield  [0]{\@secondoftwo}%
\providecommand \translation [1]{[#1]}%
\providecommand \BibitemOpen [0]{}%
\providecommand \bibitemStop [0]{}%
\providecommand \bibitemNoStop [0]{.\EOS\space}%
\providecommand \EOS [0]{\spacefactor3000\relax}%
\providecommand \BibitemShut  [1]{\csname bibitem#1\endcsname}%
\let\auto@bib@innerbib\@empty
\bibitem [{\citenamefont {Penrose}(1965)}]{Penrose:1964wq}%
  \BibitemOpen
  \bibfield  {author} {\bibinfo {author} {\bibfnamefont {R.}~\bibnamefont {Penrose}},\ }\bibfield  {title} {\bibinfo {title} {{Gravitational collapse and space-time singularities}},\ }\href {https://doi.org/10.1103/PhysRevLett.14.57} {\bibfield  {journal} {\bibinfo  {journal} {Phys. Rev. Lett.}\ }\textbf {\bibinfo {volume} {14}},\ \bibinfo {pages} {57} (\bibinfo {year} {1965})}\BibitemShut {NoStop}%
\bibitem [{\citenamefont {Hawking}(1967)}]{Hawking:1967ju}%
  \BibitemOpen
  \bibfield  {author} {\bibinfo {author} {\bibfnamefont {S.}~\bibnamefont {Hawking}},\ }\bibfield  {title} {\bibinfo {title} {{The occurrence of singularities in cosmology. III. Causality and singularities}},\ }\href {https://doi.org/10.1098/rspa.1967.0164} {\bibfield  {journal} {\bibinfo  {journal} {Proc. Roy. Soc. Lond. A}\ }\textbf {\bibinfo {volume} {300}},\ \bibinfo {pages} {187} (\bibinfo {year} {1967})}\BibitemShut {NoStop}%
\bibitem [{\citenamefont {Hawking}\ and\ \citenamefont {Penrose}(1970)}]{Hawking:1970zqf}%
  \BibitemOpen
  \bibfield  {author} {\bibinfo {author} {\bibfnamefont {S.~W.}\ \bibnamefont {Hawking}}\ and\ \bibinfo {author} {\bibfnamefont {R.}~\bibnamefont {Penrose}},\ }\bibfield  {title} {\bibinfo {title} {{The Singularities of gravitational collapse and cosmology}},\ }\href {https://doi.org/10.1098/rspa.1970.0021} {\bibfield  {journal} {\bibinfo  {journal} {Proc. Roy. Soc. Lond. A}\ }\textbf {\bibinfo {volume} {314}},\ \bibinfo {pages} {529} (\bibinfo {year} {1970})}\BibitemShut {NoStop}%
\bibitem [{\citenamefont {Gross}\ and\ \citenamefont {Sloan}(1987)}]{Gross:1986mw}%
  \BibitemOpen
  \bibfield  {author} {\bibinfo {author} {\bibfnamefont {D.~J.}\ \bibnamefont {Gross}}\ and\ \bibinfo {author} {\bibfnamefont {J.~H.}\ \bibnamefont {Sloan}},\ }\bibfield  {title} {\bibinfo {title} {{The Quartic Effective Action for the Heterotic String}},\ }\href {https://doi.org/10.1016/0550-3213(87)90465-2} {\bibfield  {journal} {\bibinfo  {journal} {Nucl. Phys. B}\ }\textbf {\bibinfo {volume} {291}},\ \bibinfo {pages} {41} (\bibinfo {year} {1987})}\BibitemShut {NoStop}%
\bibitem [{\citenamefont {Gross}\ and\ \citenamefont {Witten}(1986)}]{Gross:1986iv}%
  \BibitemOpen
  \bibfield  {author} {\bibinfo {author} {\bibfnamefont {D.~J.}\ \bibnamefont {Gross}}\ and\ \bibinfo {author} {\bibfnamefont {E.}~\bibnamefont {Witten}},\ }\bibfield  {title} {\bibinfo {title} {{Superstring Modifications of Einstein's Equations}},\ }\href {https://doi.org/10.1016/0550-3213(86)90429-3} {\bibfield  {journal} {\bibinfo  {journal} {Nucl. Phys. B}\ }\textbf {\bibinfo {volume} {277}},\ \bibinfo {pages} {1} (\bibinfo {year} {1986})}\BibitemShut {NoStop}%
\bibitem [{\citenamefont {Grisaru}\ and\ \citenamefont {Zanon}(1986)}]{Grisaru:1986vi}%
  \BibitemOpen
  \bibfield  {author} {\bibinfo {author} {\bibfnamefont {M.~T.}\ \bibnamefont {Grisaru}}\ and\ \bibinfo {author} {\bibfnamefont {D.}~\bibnamefont {Zanon}},\ }\bibfield  {title} {\bibinfo {title} {{$\sigma$ Model Superstring Corrections to the Einstein-hilbert Action}},\ }\href {https://doi.org/10.1016/0370-2693(86)90765-3} {\bibfield  {journal} {\bibinfo  {journal} {Phys. Lett. B}\ }\textbf {\bibinfo {volume} {177}},\ \bibinfo {pages} {347} (\bibinfo {year} {1986})}\BibitemShut {NoStop}%
\bibitem [{\citenamefont {Bonanno}\ \emph {et~al.}(2020)\citenamefont {Bonanno}, \citenamefont {Eichhorn}, \citenamefont {Gies}, \citenamefont {Pawlowski}, \citenamefont {Percacci}, \citenamefont {Reuter}, \citenamefont {Saueressig},\ and\ \citenamefont {Vacca}}]{Bonanno:2020bil}%
  \BibitemOpen
  \bibfield  {author} {\bibinfo {author} {\bibfnamefont {A.}~\bibnamefont {Bonanno}}, \bibinfo {author} {\bibfnamefont {A.}~\bibnamefont {Eichhorn}}, \bibinfo {author} {\bibfnamefont {H.}~\bibnamefont {Gies}}, \bibinfo {author} {\bibfnamefont {J.~M.}\ \bibnamefont {Pawlowski}}, \bibinfo {author} {\bibfnamefont {R.}~\bibnamefont {Percacci}}, \bibinfo {author} {\bibfnamefont {M.}~\bibnamefont {Reuter}}, \bibinfo {author} {\bibfnamefont {F.}~\bibnamefont {Saueressig}},\ and\ \bibinfo {author} {\bibfnamefont {G.~P.}\ \bibnamefont {Vacca}},\ }\bibfield  {title} {\bibinfo {title} {{Critical reflections on asymptotically safe gravity}},\ }\href {https://doi.org/10.3389/fphy.2020.00269} {\bibfield  {journal} {\bibinfo  {journal} {Front. in Phys.}\ }\textbf {\bibinfo {volume} {8}},\ \bibinfo {pages} {269} (\bibinfo {year} {2020})},\ \Eprint {https://arxiv.org/abs/2004.06810} {arXiv:2004.06810 [gr-qc]} \BibitemShut {NoStop}%
\bibitem [{\citenamefont {Knorr}\ \emph {et~al.}(2024)\citenamefont {Knorr}, \citenamefont {Ripken},\ and\ \citenamefont {Saueressig}}]{Knorr:2022dsx}%
  \BibitemOpen
  \bibfield  {author} {\bibinfo {author} {\bibfnamefont {B.}~\bibnamefont {Knorr}}, \bibinfo {author} {\bibfnamefont {C.}~\bibnamefont {Ripken}},\ and\ \bibinfo {author} {\bibfnamefont {F.}~\bibnamefont {Saueressig}},\ }\bibinfo {title} {{Form Factors in Asymptotically Safe Quantum Gravity}}\ (\bibinfo {year} {2024})\ \Eprint {https://arxiv.org/abs/2210.16072} {arXiv:2210.16072 [hep-th]} \BibitemShut {NoStop}%
\bibitem [{\citenamefont {Knorr}\ and\ \citenamefont {Saueressig}(2018)}]{Knorr:2018kog}%
  \BibitemOpen
  \bibfield  {author} {\bibinfo {author} {\bibfnamefont {B.}~\bibnamefont {Knorr}}\ and\ \bibinfo {author} {\bibfnamefont {F.}~\bibnamefont {Saueressig}},\ }\bibfield  {title} {\bibinfo {title} {{Towards reconstructing the quantum effective action of gravity}},\ }\href {https://doi.org/10.1103/PhysRevLett.121.161304} {\bibfield  {journal} {\bibinfo  {journal} {Phys. Rev. Lett.}\ }\textbf {\bibinfo {volume} {121}},\ \bibinfo {pages} {161304} (\bibinfo {year} {2018})},\ \Eprint {https://arxiv.org/abs/1804.03846} {arXiv:1804.03846 [hep-th]} \BibitemShut {NoStop}%
\bibitem [{\citenamefont {Dupuis}\ \emph {et~al.}(2021)\citenamefont {Dupuis}, \citenamefont {Canet}, \citenamefont {Eichhorn}, \citenamefont {Metzner}, \citenamefont {Pawlowski}, \citenamefont {Tissier},\ and\ \citenamefont {Wschebor}}]{Dupuis:2020fhh}%
  \BibitemOpen
  \bibfield  {author} {\bibinfo {author} {\bibfnamefont {N.}~\bibnamefont {Dupuis}}, \bibinfo {author} {\bibfnamefont {L.}~\bibnamefont {Canet}}, \bibinfo {author} {\bibfnamefont {A.}~\bibnamefont {Eichhorn}}, \bibinfo {author} {\bibfnamefont {W.}~\bibnamefont {Metzner}}, \bibinfo {author} {\bibfnamefont {J.~M.}\ \bibnamefont {Pawlowski}}, \bibinfo {author} {\bibfnamefont {M.}~\bibnamefont {Tissier}},\ and\ \bibinfo {author} {\bibfnamefont {N.}~\bibnamefont {Wschebor}},\ }\bibfield  {title} {\bibinfo {title} {{The nonperturbative functional renormalization group and its applications}},\ }\href {https://doi.org/10.1016/j.physrep.2021.01.001} {\bibfield  {journal} {\bibinfo  {journal} {Phys. Rept.}\ }\textbf {\bibinfo {volume} {910}},\ \bibinfo {pages} {1} (\bibinfo {year} {2021})},\ \Eprint {https://arxiv.org/abs/2006.04853} {arXiv:2006.04853 [cond-mat.stat-mech]} \BibitemShut {NoStop}%
\bibitem [{\citenamefont {Bueno}\ \emph {et~al.}(2025{\natexlab{a}})\citenamefont {Bueno}, \citenamefont {Cano},\ and\ \citenamefont {Hennigar}}]{Bueno:2024dgm}%
  \BibitemOpen
  \bibfield  {author} {\bibinfo {author} {\bibfnamefont {P.}~\bibnamefont {Bueno}}, \bibinfo {author} {\bibfnamefont {P.~A.}\ \bibnamefont {Cano}},\ and\ \bibinfo {author} {\bibfnamefont {R.~A.}\ \bibnamefont {Hennigar}},\ }\bibfield  {title} {\bibinfo {title} {{Regular black holes from pure gravity}},\ }\href {https://doi.org/10.1016/j.physletb.2025.139260} {\bibfield  {journal} {\bibinfo  {journal} {Phys. Lett. B}\ }\textbf {\bibinfo {volume} {861}},\ \bibinfo {pages} {139260} (\bibinfo {year} {2025}{\natexlab{a}})},\ \Eprint {https://arxiv.org/abs/2403.04827} {arXiv:2403.04827 [gr-qc]} \BibitemShut {NoStop}%
\bibitem [{\citenamefont {Bueno}\ \emph {et~al.}(2024{\natexlab{a}})\citenamefont {Bueno}, \citenamefont {Cano}, \citenamefont {Hennigar},\ and\ \citenamefont {Murcia}}]{Bueno:2024zsx}%
  \BibitemOpen
  \bibfield  {author} {\bibinfo {author} {\bibfnamefont {P.}~\bibnamefont {Bueno}}, \bibinfo {author} {\bibfnamefont {P.~A.}\ \bibnamefont {Cano}}, \bibinfo {author} {\bibfnamefont {R.~A.}\ \bibnamefont {Hennigar}},\ and\ \bibinfo {author} {\bibfnamefont {A.~J.}\ \bibnamefont {Murcia}},\ }\href@noop {} {\bibinfo {title} {{Regular black holes from thin-shell collapse}}} (\bibinfo {year} {2024}{\natexlab{a}}),\ \Eprint {https://arxiv.org/abs/2412.02740} {arXiv:2412.02740 [gr-qc]} \BibitemShut {NoStop}%
\bibitem [{\citenamefont {Bueno}\ \emph {et~al.}(2024{\natexlab{b}})\citenamefont {Bueno}, \citenamefont {Cano}, \citenamefont {Hennigar},\ and\ \citenamefont {Murcia}}]{Bueno:2024eig}%
  \BibitemOpen
  \bibfield  {author} {\bibinfo {author} {\bibfnamefont {P.}~\bibnamefont {Bueno}}, \bibinfo {author} {\bibfnamefont {P.~A.}\ \bibnamefont {Cano}}, \bibinfo {author} {\bibfnamefont {R.~A.}\ \bibnamefont {Hennigar}},\ and\ \bibinfo {author} {\bibfnamefont {A.~J.}\ \bibnamefont {Murcia}},\ }\href@noop {} {\bibinfo {title} {{Dynamical Formation of Regular Black Holes}}} (\bibinfo {year} {2024}{\natexlab{b}}),\ \Eprint {https://arxiv.org/abs/2412.02742} {arXiv:2412.02742 [gr-qc]} \BibitemShut {NoStop}%
\bibitem [{\citenamefont {{Aguilar-Gutierrez}}\ \emph {et~al.}(2023)\citenamefont {{Aguilar-Gutierrez}}, \citenamefont {Bueno}, \citenamefont {Cano}, \citenamefont {Hennigar},\ and\ \citenamefont {Llorens}}]{aguilar-gutierrezAspectsHighercurvatureGravities2023}%
  \BibitemOpen
  \bibfield  {author} {\bibinfo {author} {\bibfnamefont {S.~E.}\ \bibnamefont {{Aguilar-Gutierrez}}}, \bibinfo {author} {\bibfnamefont {P.}~\bibnamefont {Bueno}}, \bibinfo {author} {\bibfnamefont {P.~A.}\ \bibnamefont {Cano}}, \bibinfo {author} {\bibfnamefont {R.~A.}\ \bibnamefont {Hennigar}},\ and\ \bibinfo {author} {\bibfnamefont {Q.}~\bibnamefont {Llorens}},\ }\href {https://doi.org/10.48550/arXiv.2310.09333} {\bibinfo {title} {Aspects of higher-curvature gravities with covariant derivatives}} (\bibinfo {year} {2023}),\ \Eprint {https://arxiv.org/abs/2310.09333} {arXiv:2310.09333} \BibitemShut {NoStop}%
\bibitem [{\citenamefont {Ahmed}\ \emph {et~al.}(2017)\citenamefont {Ahmed}, \citenamefont {Hennigar}, \citenamefont {Mann},\ and\ \citenamefont {Mir}}]{ahmedQuintessentialQuarticQuasitopological2017}%
  \BibitemOpen
  \bibfield  {author} {\bibinfo {author} {\bibfnamefont {J.}~\bibnamefont {Ahmed}}, \bibinfo {author} {\bibfnamefont {R.~A.}\ \bibnamefont {Hennigar}}, \bibinfo {author} {\bibfnamefont {R.~B.}\ \bibnamefont {Mann}},\ and\ \bibinfo {author} {\bibfnamefont {M.}~\bibnamefont {Mir}},\ }\href {https://doi.org/10.48550/arXiv.1703.11007} {\bibinfo {title} {Quintessential {{Quartic Quasi-topological Quartet}}}} (\bibinfo {year} {2017}),\ \Eprint {https://arxiv.org/abs/1703.11007} {arXiv:1703.11007} \BibitemShut {NoStop}%
\bibitem [{\citenamefont {Bueno}\ and\ \citenamefont {Cano}(2017{\natexlab{a}})}]{buenoFourdimensionalBlackHoles2017}%
  \BibitemOpen
  \bibfield  {author} {\bibinfo {author} {\bibfnamefont {P.}~\bibnamefont {Bueno}}\ and\ \bibinfo {author} {\bibfnamefont {P.~A.}\ \bibnamefont {Cano}},\ }\href {https://doi.org/10.48550/arXiv.1610.08019} {\bibinfo {title} {Four-dimensional black holes in {{Einsteinian}} cubic gravity}} (\bibinfo {year} {2017}{\natexlab{a}}),\ \Eprint {https://arxiv.org/abs/1610.08019} {arXiv:1610.08019} \BibitemShut {NoStop}%
\bibitem [{\citenamefont {Bueno}\ \emph {et~al.}(2019)\citenamefont {Bueno}, \citenamefont {Cano},\ and\ \citenamefont {Hennigar}}]{buenoGeneralizedQuasitopologicalGravities2019}%
  \BibitemOpen
  \bibfield  {author} {\bibinfo {author} {\bibfnamefont {P.}~\bibnamefont {Bueno}}, \bibinfo {author} {\bibfnamefont {P.~A.}\ \bibnamefont {Cano}},\ and\ \bibinfo {author} {\bibfnamefont {R.~A.}\ \bibnamefont {Hennigar}},\ }\href {https://doi.org/10.48550/arXiv.1909.07983} {\bibinfo {title} {({{Generalized}}) quasi-topological gravities at all orders}} (\bibinfo {year} {2019}),\ \Eprint {https://arxiv.org/abs/1909.07983} {arXiv:1909.07983} \BibitemShut {NoStop}%
\bibitem [{\citenamefont {Bueno}\ \emph {et~al.}(2022)\citenamefont {Bueno}, \citenamefont {Cano}, \citenamefont {Hennigar}, \citenamefont {Lu},\ and\ \citenamefont {Moreno}}]{buenoGeneralizedQuasitopologicalGravities2022}%
  \BibitemOpen
  \bibfield  {author} {\bibinfo {author} {\bibfnamefont {P.}~\bibnamefont {Bueno}}, \bibinfo {author} {\bibfnamefont {P.~A.}\ \bibnamefont {Cano}}, \bibinfo {author} {\bibfnamefont {R.~A.}\ \bibnamefont {Hennigar}}, \bibinfo {author} {\bibfnamefont {M.}~\bibnamefont {Lu}},\ and\ \bibinfo {author} {\bibfnamefont {J.}~\bibnamefont {Moreno}},\ }\href {https://doi.org/10.48550/arXiv.2203.05589} {\bibinfo {title} {Generalized quasi-topological gravities: The whole shebang}} (\bibinfo {year} {2022}),\ \Eprint {https://arxiv.org/abs/2203.05589} {arXiv:2203.05589} \BibitemShut {NoStop}%
\bibitem [{\citenamefont {Bueno}\ and\ \citenamefont {Cano}(2017{\natexlab{b}})}]{buenoUniversalBlackHole2017}%
  \BibitemOpen
  \bibfield  {author} {\bibinfo {author} {\bibfnamefont {P.}~\bibnamefont {Bueno}}\ and\ \bibinfo {author} {\bibfnamefont {P.~A.}\ \bibnamefont {Cano}},\ }\href {https://doi.org/10.48550/arXiv.1704.02967} {\bibinfo {title} {Universal black hole stability in four dimensions}} (\bibinfo {year} {2017}{\natexlab{b}}),\ \Eprint {https://arxiv.org/abs/1704.02967} {arXiv:1704.02967} \BibitemShut {NoStop}%
\bibitem [{\citenamefont {Hennigar}\ and\ \citenamefont {Mann}(2016)}]{hennigarBlackHolesEinsteinian2016}%
  \BibitemOpen
  \bibfield  {author} {\bibinfo {author} {\bibfnamefont {R.~A.}\ \bibnamefont {Hennigar}}\ and\ \bibinfo {author} {\bibfnamefont {R.~B.}\ \bibnamefont {Mann}},\ }\href {https://doi.org/10.48550/arXiv.1610.06675} {\bibinfo {title} {Black holes in {{Einsteinian}} cubic gravity}} (\bibinfo {year} {2016}),\ \Eprint {https://arxiv.org/abs/1610.06675} {arXiv:1610.06675} \BibitemShut {NoStop}%
\bibitem [{\citenamefont {Hennigar}\ \emph {et~al.}(2017)\citenamefont {Hennigar}, \citenamefont {Kubiznak},\ and\ \citenamefont {Mann}}]{hennigarGeneralizedQuasitopologicalGravity2017}%
  \BibitemOpen
  \bibfield  {author} {\bibinfo {author} {\bibfnamefont {R.~A.}\ \bibnamefont {Hennigar}}, \bibinfo {author} {\bibfnamefont {D.}~\bibnamefont {Kubiznak}},\ and\ \bibinfo {author} {\bibfnamefont {R.~B.}\ \bibnamefont {Mann}},\ }\href {https://doi.org/10.48550/arXiv.1703.01631} {\bibinfo {title} {Generalized quasi-topological gravity}} (\bibinfo {year} {2017}),\ \Eprint {https://arxiv.org/abs/1703.01631} {arXiv:1703.01631} \BibitemShut {NoStop}%
\bibitem [{\citenamefont {Moreno}\ and\ \citenamefont {Murcia}(2023)}]{morenoClassificationGeneralizedQuasitopological2023a}%
  \BibitemOpen
  \bibfield  {author} {\bibinfo {author} {\bibfnamefont {J.}~\bibnamefont {Moreno}}\ and\ \bibinfo {author} {\bibfnamefont {{\'A}.~J.}\ \bibnamefont {Murcia}},\ }\href {https://doi.org/10.48550/arXiv.2304.08510} {\bibinfo {title} {On the classification of {{Generalized Quasitopological Gravities}}}} (\bibinfo {year} {2023}),\ \Eprint {https://arxiv.org/abs/2304.08510} {arXiv:2304.08510} \BibitemShut {NoStop}%
\bibitem [{\citenamefont {Moreno}\ and\ \citenamefont {Murcia}(2024)}]{Moreno:2023arp}%
  \BibitemOpen
  \bibfield  {author} {\bibinfo {author} {\bibfnamefont {J.}~\bibnamefont {Moreno}}\ and\ \bibinfo {author} {\bibfnamefont {A.~J.}\ \bibnamefont {Murcia}},\ }\bibfield  {title} {\bibinfo {title} {{Cosmological higher-curvature gravities}},\ }\href {https://doi.org/10.1088/1361-6382/ad51c5} {\bibfield  {journal} {\bibinfo  {journal} {Class. Quant. Grav.}\ }\textbf {\bibinfo {volume} {41}},\ \bibinfo {pages} {135017} (\bibinfo {year} {2024})},\ \Eprint {https://arxiv.org/abs/2311.12104} {arXiv:2311.12104 [gr-qc]} \BibitemShut {NoStop}%
\bibitem [{\citenamefont {Cisterna}\ \emph {et~al.}(2020)\citenamefont {Cisterna}, \citenamefont {Grandi},\ and\ \citenamefont {Oliva}}]{Cisterna:2018tgx}%
  \BibitemOpen
  \bibfield  {author} {\bibinfo {author} {\bibfnamefont {A.}~\bibnamefont {Cisterna}}, \bibinfo {author} {\bibfnamefont {N.}~\bibnamefont {Grandi}},\ and\ \bibinfo {author} {\bibfnamefont {J.}~\bibnamefont {Oliva}},\ }\bibfield  {title} {\bibinfo {title} {{On four-dimensional Einsteinian gravity, quasitopological gravity, cosmology and black holes}},\ }\href {https://doi.org/10.1016/j.physletb.2020.135435} {\bibfield  {journal} {\bibinfo  {journal} {Phys. Lett. B}\ }\textbf {\bibinfo {volume} {805}},\ \bibinfo {pages} {135435} (\bibinfo {year} {2020})},\ \Eprint {https://arxiv.org/abs/1811.06523} {arXiv:1811.06523 [hep-th]} \BibitemShut {NoStop}%
\bibitem [{\citenamefont {Carballo-Rubio}\ \emph {et~al.}(2025)\citenamefont {Carballo-Rubio} \emph {et~al.}}]{Carballo-Rubio:2025fnc}%
  \BibitemOpen
  \bibfield  {author} {\bibinfo {author} {\bibfnamefont {R.}~\bibnamefont {Carballo-Rubio}} \emph {et~al.},\ }\href@noop {} {\bibinfo {title} {{Towards a Non-singular Paradigm of Black Hole Physics}}} (\bibinfo {year} {2025}),\ \Eprint {https://arxiv.org/abs/2501.05505} {arXiv:2501.05505 [gr-qc]} \BibitemShut {NoStop}%
\bibitem [{\citenamefont {Ayon-Beato}\ and\ \citenamefont {Garcia}(1998)}]{Ayon-Beato:1998hmi}%
  \BibitemOpen
  \bibfield  {author} {\bibinfo {author} {\bibfnamefont {E.}~\bibnamefont {Ayon-Beato}}\ and\ \bibinfo {author} {\bibfnamefont {A.}~\bibnamefont {Garcia}},\ }\bibfield  {title} {\bibinfo {title} {{Regular black hole in general relativity coupled to nonlinear electrodynamics}},\ }\href {https://doi.org/10.1103/PhysRevLett.80.5056} {\bibfield  {journal} {\bibinfo  {journal} {Phys. Rev. Lett.}\ }\textbf {\bibinfo {volume} {80}},\ \bibinfo {pages} {5056} (\bibinfo {year} {1998})},\ \Eprint {https://arxiv.org/abs/gr-qc/9911046} {arXiv:gr-qc/9911046} \BibitemShut {NoStop}%
\bibitem [{\citenamefont {Ayon-Beato}\ and\ \citenamefont {Garcia}(2000)}]{Ayon-Beato:2000mjt}%
  \BibitemOpen
  \bibfield  {author} {\bibinfo {author} {\bibfnamefont {E.}~\bibnamefont {Ayon-Beato}}\ and\ \bibinfo {author} {\bibfnamefont {A.}~\bibnamefont {Garcia}},\ }\bibfield  {title} {\bibinfo {title} {{The Bardeen model as a nonlinear magnetic monopole}},\ }\href {https://doi.org/10.1016/S0370-2693(00)01125-4} {\bibfield  {journal} {\bibinfo  {journal} {Phys. Lett. B}\ }\textbf {\bibinfo {volume} {493}},\ \bibinfo {pages} {149} (\bibinfo {year} {2000})},\ \Eprint {https://arxiv.org/abs/gr-qc/0009077} {arXiv:gr-qc/0009077} \BibitemShut {NoStop}%
\bibitem [{\citenamefont {Bronnikov}(2022)}]{Bronnikov:2022ofk}%
  \BibitemOpen
  \bibfield  {author} {\bibinfo {author} {\bibfnamefont {K.~A.}\ \bibnamefont {Bronnikov}},\ }\href@noop {} {\bibinfo {title} {{Regular black holes sourced by nonlinear electrodynamics}}} (\bibinfo {year} {2022}),\ \Eprint {https://arxiv.org/abs/2211.00743} {arXiv:2211.00743 [gr-qc]} \BibitemShut {NoStop}%
\bibitem [{\citenamefont {Arciniega}\ \emph {et~al.}(2020{\natexlab{a}})\citenamefont {Arciniega}, \citenamefont {Bueno}, \citenamefont {Cano}, \citenamefont {Edelstein}, \citenamefont {Hennigar},\ and\ \citenamefont {Jaime}}]{Arciniega:2018tnn}%
  \BibitemOpen
  \bibfield  {author} {\bibinfo {author} {\bibfnamefont {G.}~\bibnamefont {Arciniega}}, \bibinfo {author} {\bibfnamefont {P.}~\bibnamefont {Bueno}}, \bibinfo {author} {\bibfnamefont {P.~A.}\ \bibnamefont {Cano}}, \bibinfo {author} {\bibfnamefont {J.~D.}\ \bibnamefont {Edelstein}}, \bibinfo {author} {\bibfnamefont {R.~A.}\ \bibnamefont {Hennigar}},\ and\ \bibinfo {author} {\bibfnamefont {L.~G.}\ \bibnamefont {Jaime}},\ }\bibfield  {title} {\bibinfo {title} {{Geometric Inflation}},\ }\href {https://doi.org/10.1016/j.physletb.2020.135242} {\bibfield  {journal} {\bibinfo  {journal} {Phys. Lett. B}\ }\textbf {\bibinfo {volume} {802}},\ \bibinfo {pages} {135242} (\bibinfo {year} {2020}{\natexlab{a}})},\ \Eprint {https://arxiv.org/abs/1812.11187} {arXiv:1812.11187 [hep-th]} \BibitemShut {NoStop}%
\bibitem [{\citenamefont {Arciniega}\ \emph {et~al.}(2020{\natexlab{b}})\citenamefont {Arciniega}, \citenamefont {Edelstein},\ and\ \citenamefont {Jaime}}]{Arciniega:2018fxj}%
  \BibitemOpen
  \bibfield  {author} {\bibinfo {author} {\bibfnamefont {G.}~\bibnamefont {Arciniega}}, \bibinfo {author} {\bibfnamefont {J.~D.}\ \bibnamefont {Edelstein}},\ and\ \bibinfo {author} {\bibfnamefont {L.~G.}\ \bibnamefont {Jaime}},\ }\bibfield  {title} {\bibinfo {title} {{Towards geometric inflation: the cubic case}},\ }\href {https://doi.org/10.1016/j.physletb.2020.135272} {\bibfield  {journal} {\bibinfo  {journal} {Phys. Lett. B}\ }\textbf {\bibinfo {volume} {802}},\ \bibinfo {pages} {135272} (\bibinfo {year} {2020}{\natexlab{b}})},\ \Eprint {https://arxiv.org/abs/1810.08166} {arXiv:1810.08166 [gr-qc]} \BibitemShut {NoStop}%
\bibitem [{\citenamefont {Arciniega}\ \emph {et~al.}(2020{\natexlab{c}})\citenamefont {Arciniega}, \citenamefont {Jaime},\ and\ \citenamefont {Piccinelli}}]{Arciniega:2020pcy}%
  \BibitemOpen
  \bibfield  {author} {\bibinfo {author} {\bibfnamefont {G.}~\bibnamefont {Arciniega}}, \bibinfo {author} {\bibfnamefont {L.}~\bibnamefont {Jaime}},\ and\ \bibinfo {author} {\bibfnamefont {G.}~\bibnamefont {Piccinelli}},\ }\bibfield  {title} {\bibinfo {title} {{Inflationary predictions of Geometric Inflation}},\ }\href {https://doi.org/10.1016/j.physletb.2020.135731} {\bibfield  {journal} {\bibinfo  {journal} {Phys. Lett. B}\ }\textbf {\bibinfo {volume} {809}},\ \bibinfo {pages} {135731} (\bibinfo {year} {2020}{\natexlab{c}})},\ \Eprint {https://arxiv.org/abs/2001.11094} {arXiv:2001.11094 [gr-qc]} \BibitemShut {NoStop}%
\bibitem [{\citenamefont {Edelstein}\ \emph {et~al.}(2020)\citenamefont {Edelstein}, \citenamefont {V\'azquez~Rodr\'\i{}guez},\ and\ \citenamefont {Vilar~L\'opez}}]{Edelstein:2020nhg}%
  \BibitemOpen
  \bibfield  {author} {\bibinfo {author} {\bibfnamefont {J.~D.}\ \bibnamefont {Edelstein}}, \bibinfo {author} {\bibfnamefont {D.}~\bibnamefont {V\'azquez~Rodr\'\i{}guez}},\ and\ \bibinfo {author} {\bibfnamefont {A.}~\bibnamefont {Vilar~L\'opez}},\ }\bibfield  {title} {\bibinfo {title} {{Aspects of Geometric Inflation}},\ }\href {https://doi.org/10.1088/1475-7516/2020/12/040} {\bibfield  {journal} {\bibinfo  {journal} {JCAP}\ }\textbf {\bibinfo {volume} {12}},\ \bibinfo {pages} {040}},\ \Eprint {https://arxiv.org/abs/2006.10007} {arXiv:2006.10007 [hep-th]} \BibitemShut {NoStop}%
\bibitem [{\citenamefont {Jaime}(2021)}]{Jaime:2021pqn}%
  \BibitemOpen
  \bibfield  {author} {\bibinfo {author} {\bibfnamefont {L.~G.}\ \bibnamefont {Jaime}},\ }\bibfield  {title} {\bibinfo {title} {{On the viability of the evolution of the universe with Geometric Inflation}},\ }\href {https://doi.org/10.1016/j.dark.2021.100887} {\bibfield  {journal} {\bibinfo  {journal} {Phys. Dark Univ.}\ }\textbf {\bibinfo {volume} {34}},\ \bibinfo {pages} {100887} (\bibinfo {year} {2021})},\ \Eprint {https://arxiv.org/abs/2109.11681} {arXiv:2109.11681 [gr-qc]} \BibitemShut {NoStop}%
\bibitem [{\citenamefont {Arciniega}\ \emph {et~al.}(2025)\citenamefont {Arciniega}, \citenamefont {Jaime}, \citenamefont {Landau},\ and\ \citenamefont {Leizerovich}}]{Arciniega:2025viq}%
  \BibitemOpen
  \bibfield  {author} {\bibinfo {author} {\bibfnamefont {G.}~\bibnamefont {Arciniega}}, \bibinfo {author} {\bibfnamefont {L.~G.}\ \bibnamefont {Jaime}}, \bibinfo {author} {\bibfnamefont {S.~J.}\ \bibnamefont {Landau}},\ and\ \bibinfo {author} {\bibfnamefont {M.}~\bibnamefont {Leizerovich}},\ }\href@noop {} {\bibinfo {title} {{On Geometric Cosmology}}} (\bibinfo {year} {2025}),\ \Eprint {https://arxiv.org/abs/2504.00124} {arXiv:2504.00124 [gr-qc]} \BibitemShut {NoStop}%
\bibitem [{\citenamefont {Bojowald}(2002)}]{Bojowald:2002nz}%
  \BibitemOpen
  \bibfield  {author} {\bibinfo {author} {\bibfnamefont {M.}~\bibnamefont {Bojowald}},\ }\bibfield  {title} {\bibinfo {title} {{Inflation from quantum geometry}},\ }\href {https://doi.org/10.1103/PhysRevLett.89.261301} {\bibfield  {journal} {\bibinfo  {journal} {Phys. Rev. Lett.}\ }\textbf {\bibinfo {volume} {89}},\ \bibinfo {pages} {261301} (\bibinfo {year} {2002})},\ \Eprint {https://arxiv.org/abs/gr-qc/0206054} {arXiv:gr-qc/0206054} \BibitemShut {NoStop}%
\bibitem [{\citenamefont {de~Cesare}\ and\ \citenamefont {Sakellariadou}(2017)}]{deCesare:2016axk}%
  \BibitemOpen
  \bibfield  {author} {\bibinfo {author} {\bibfnamefont {M.}~\bibnamefont {de~Cesare}}\ and\ \bibinfo {author} {\bibfnamefont {M.}~\bibnamefont {Sakellariadou}},\ }\bibfield  {title} {\bibinfo {title} {{Accelerated expansion of the Universe without an inflaton and resolution of the initial singularity from Group Field Theory condensates}},\ }\href {https://doi.org/10.1016/j.physletb.2016.10.051} {\bibfield  {journal} {\bibinfo  {journal} {Phys. Lett. B}\ }\textbf {\bibinfo {volume} {764}},\ \bibinfo {pages} {49} (\bibinfo {year} {2017})},\ \Eprint {https://arxiv.org/abs/1603.01764} {arXiv:1603.01764 [gr-qc]} \BibitemShut {NoStop}%
\bibitem [{\citenamefont {Starobinsky}(1980)}]{Starobinsky:1980te}%
  \BibitemOpen
  \bibfield  {author} {\bibinfo {author} {\bibfnamefont {A.~A.}\ \bibnamefont {Starobinsky}},\ }\bibfield  {title} {\bibinfo {title} {{A New Type of Isotropic Cosmological Models Without Singularity}},\ }\href {https://doi.org/10.1016/0370-2693(80)90670-X} {\bibfield  {journal} {\bibinfo  {journal} {Phys. Lett. B}\ }\textbf {\bibinfo {volume} {91}},\ \bibinfo {pages} {99} (\bibinfo {year} {1980})}\BibitemShut {NoStop}%
\bibitem [{\citenamefont {Beltr\'an~Jim\'enez}\ and\ \citenamefont {Jim\'enez-Cano}(2024)}]{BeltranJimenez:2023mxp}%
  \BibitemOpen
  \bibfield  {author} {\bibinfo {author} {\bibfnamefont {J.}~\bibnamefont {Beltr\'an~Jim\'enez}}\ and\ \bibinfo {author} {\bibfnamefont {A.}~\bibnamefont {Jim\'enez-Cano}},\ }\bibfield  {title} {\bibinfo {title} {{On the physical viability of black hole solutions in Einsteinian Cubic Gravity and its generalisations}},\ }\href {https://doi.org/10.1016/j.dark.2023.101387} {\bibfield  {journal} {\bibinfo  {journal} {Phys. Dark Univ.}\ }\textbf {\bibinfo {volume} {43}},\ \bibinfo {pages} {101387} (\bibinfo {year} {2024})},\ \Eprint {https://arxiv.org/abs/2306.07095} {arXiv:2306.07095 [gr-qc]} \BibitemShut {NoStop}%
\bibitem [{\citenamefont {De~Felice}\ and\ \citenamefont {Tsujikawa}(2023)}]{DeFelice:2023vmj}%
  \BibitemOpen
  \bibfield  {author} {\bibinfo {author} {\bibfnamefont {A.}~\bibnamefont {De~Felice}}\ and\ \bibinfo {author} {\bibfnamefont {S.}~\bibnamefont {Tsujikawa}},\ }\bibfield  {title} {\bibinfo {title} {{Excluding static and spherically symmetric black holes in Einsteinian cubic gravity with unsuppressed higher-order curvature terms}},\ }\href {https://doi.org/10.1016/j.physletb.2023.138047} {\bibfield  {journal} {\bibinfo  {journal} {Phys. Lett. B}\ }\textbf {\bibinfo {volume} {843}},\ \bibinfo {pages} {138047} (\bibinfo {year} {2023})},\ \Eprint {https://arxiv.org/abs/2305.07217} {arXiv:2305.07217 [gr-qc]} \BibitemShut {NoStop}%
\bibitem [{\citenamefont {Lovelock}(1971)}]{Lovelock:1971yv}%
  \BibitemOpen
  \bibfield  {author} {\bibinfo {author} {\bibfnamefont {D.}~\bibnamefont {Lovelock}},\ }\bibfield  {title} {\bibinfo {title} {{The Einstein tensor and its generalizations}},\ }\href {https://doi.org/10.1063/1.1665613} {\bibfield  {journal} {\bibinfo  {journal} {J. Math. Phys.}\ }\textbf {\bibinfo {volume} {12}},\ \bibinfo {pages} {498} (\bibinfo {year} {1971})}\BibitemShut {NoStop}%
\bibitem [{\citenamefont {Padmanabhan}\ and\ \citenamefont {Kothawala}(2013)}]{Padmanabhan:2013xyr}%
  \BibitemOpen
  \bibfield  {author} {\bibinfo {author} {\bibfnamefont {T.}~\bibnamefont {Padmanabhan}}\ and\ \bibinfo {author} {\bibfnamefont {D.}~\bibnamefont {Kothawala}},\ }\bibfield  {title} {\bibinfo {title} {{Lanczos-Lovelock models of gravity}},\ }\href {https://doi.org/10.1016/j.physrep.2013.05.007} {\bibfield  {journal} {\bibinfo  {journal} {Phys. Rept.}\ }\textbf {\bibinfo {volume} {531}},\ \bibinfo {pages} {115} (\bibinfo {year} {2013})},\ \Eprint {https://arxiv.org/abs/1302.2151} {arXiv:1302.2151 [gr-qc]} \BibitemShut {NoStop}%
\bibitem [{\citenamefont {Glavan}\ and\ \citenamefont {Lin}(2020)}]{Glavan:2019inb}%
  \BibitemOpen
  \bibfield  {author} {\bibinfo {author} {\bibfnamefont {D.}~\bibnamefont {Glavan}}\ and\ \bibinfo {author} {\bibfnamefont {C.}~\bibnamefont {Lin}},\ }\bibfield  {title} {\bibinfo {title} {{Einstein-Gauss-Bonnet Gravity in Four-Dimensional Spacetime}},\ }\href {https://doi.org/10.1103/PhysRevLett.124.081301} {\bibfield  {journal} {\bibinfo  {journal} {Phys. Rev. Lett.}\ }\textbf {\bibinfo {volume} {124}},\ \bibinfo {pages} {081301} (\bibinfo {year} {2020})},\ \Eprint {https://arxiv.org/abs/1905.03601} {arXiv:1905.03601 [gr-qc]} \BibitemShut {NoStop}%
\bibitem [{\citenamefont {G\"urses}\ \emph {et~al.}(2020)\citenamefont {G\"urses}, \citenamefont {\c{S}i\c{s}man},\ and\ \citenamefont {Tekin}}]{Gurses:2020ofy}%
  \BibitemOpen
  \bibfield  {author} {\bibinfo {author} {\bibfnamefont {M.}~\bibnamefont {G\"urses}}, \bibinfo {author} {\bibfnamefont {T.~c.}\ \bibnamefont {\c{S}i\c{s}man}},\ and\ \bibinfo {author} {\bibfnamefont {B.}~\bibnamefont {Tekin}},\ }\bibfield  {title} {\bibinfo {title} {{Is there a novel Einstein\textendash{}Gauss\textendash{}Bonnet theory in four dimensions?}},\ }\href {https://doi.org/10.1140/epjc/s10052-020-8200-7} {\bibfield  {journal} {\bibinfo  {journal} {Eur. Phys. J. C}\ }\textbf {\bibinfo {volume} {80}},\ \bibinfo {pages} {647} (\bibinfo {year} {2020})},\ \Eprint {https://arxiv.org/abs/2004.03390} {arXiv:2004.03390 [gr-qc]} \BibitemShut {NoStop}%
\bibitem [{\citenamefont {Fernandes}\ \emph {et~al.}(2020)\citenamefont {Fernandes}, \citenamefont {Carrilho}, \citenamefont {Clifton},\ and\ \citenamefont {Mulryne}}]{Fernandes:2020nbq}%
  \BibitemOpen
  \bibfield  {author} {\bibinfo {author} {\bibfnamefont {P.~G.~S.}\ \bibnamefont {Fernandes}}, \bibinfo {author} {\bibfnamefont {P.}~\bibnamefont {Carrilho}}, \bibinfo {author} {\bibfnamefont {T.}~\bibnamefont {Clifton}},\ and\ \bibinfo {author} {\bibfnamefont {D.~J.}\ \bibnamefont {Mulryne}},\ }\bibfield  {title} {\bibinfo {title} {{Derivation of Regularized Field Equations for the Einstein-Gauss-Bonnet Theory in Four Dimensions}},\ }\href {https://doi.org/10.1103/PhysRevD.102.024025} {\bibfield  {journal} {\bibinfo  {journal} {Phys. Rev. D}\ }\textbf {\bibinfo {volume} {102}},\ \bibinfo {pages} {024025} (\bibinfo {year} {2020})},\ \Eprint {https://arxiv.org/abs/2004.08362} {arXiv:2004.08362 [gr-qc]} \BibitemShut {NoStop}%
\bibitem [{\citenamefont {Hennigar}\ \emph {et~al.}(2020{\natexlab{a}})\citenamefont {Hennigar}, \citenamefont {Kubiz\v{n}\'ak}, \citenamefont {Mann},\ and\ \citenamefont {Pollack}}]{Hennigar:2020lsl}%
  \BibitemOpen
  \bibfield  {author} {\bibinfo {author} {\bibfnamefont {R.~A.}\ \bibnamefont {Hennigar}}, \bibinfo {author} {\bibfnamefont {D.}~\bibnamefont {Kubiz\v{n}\'ak}}, \bibinfo {author} {\bibfnamefont {R.~B.}\ \bibnamefont {Mann}},\ and\ \bibinfo {author} {\bibfnamefont {C.}~\bibnamefont {Pollack}},\ }\bibfield  {title} {\bibinfo {title} {{On taking the D \textrightarrow{} 4 limit of Gauss-Bonnet gravity: theory and solutions}},\ }\href {https://doi.org/10.1007/JHEP07(2020)027} {\bibfield  {journal} {\bibinfo  {journal} {JHEP}\ }\textbf {\bibinfo {volume} {07}},\ \bibinfo {pages} {027}},\ \Eprint {https://arxiv.org/abs/2004.09472} {arXiv:2004.09472 [gr-qc]} \BibitemShut {NoStop}%
\bibitem [{\citenamefont {Lu}\ and\ \citenamefont {Pang}(2020)}]{Lu:2020iav}%
  \BibitemOpen
  \bibfield  {author} {\bibinfo {author} {\bibfnamefont {H.}~\bibnamefont {Lu}}\ and\ \bibinfo {author} {\bibfnamefont {Y.}~\bibnamefont {Pang}},\ }\bibfield  {title} {\bibinfo {title} {{Horndeski gravity as $D \rightarrow 4$ limit of Gauss-Bonnet}},\ }\href {https://doi.org/10.1016/j.physletb.2020.135717} {\bibfield  {journal} {\bibinfo  {journal} {Phys. Lett. B}\ }\textbf {\bibinfo {volume} {809}},\ \bibinfo {pages} {135717} (\bibinfo {year} {2020})},\ \Eprint {https://arxiv.org/abs/2003.11552} {arXiv:2003.11552 [gr-qc]} \BibitemShut {NoStop}%
\bibitem [{\citenamefont {Kobayashi}(2020)}]{Kobayashi:2020wqy}%
  \BibitemOpen
  \bibfield  {author} {\bibinfo {author} {\bibfnamefont {T.}~\bibnamefont {Kobayashi}},\ }\bibfield  {title} {\bibinfo {title} {{Effective scalar-tensor description of regularized Lovelock gravity in four dimensions}},\ }\href {https://doi.org/10.1088/1475-7516/2020/07/013} {\bibfield  {journal} {\bibinfo  {journal} {JCAP}\ }\textbf {\bibinfo {volume} {07}},\ \bibinfo {pages} {013}},\ \Eprint {https://arxiv.org/abs/2003.12771} {arXiv:2003.12771 [gr-qc]} \BibitemShut {NoStop}%
\bibitem [{\citenamefont {Fernandes}(2021)}]{Fernandes:2021dsb}%
  \BibitemOpen
  \bibfield  {author} {\bibinfo {author} {\bibfnamefont {P.~G.~S.}\ \bibnamefont {Fernandes}},\ }\bibfield  {title} {\bibinfo {title} {{Gravity with a generalized conformal scalar field: theory and solutions}},\ }\href {https://doi.org/10.1103/PhysRevD.103.104065} {\bibfield  {journal} {\bibinfo  {journal} {Phys. Rev. D}\ }\textbf {\bibinfo {volume} {103}},\ \bibinfo {pages} {104065} (\bibinfo {year} {2021})},\ \Eprint {https://arxiv.org/abs/2105.04687} {arXiv:2105.04687 [gr-qc]} \BibitemShut {NoStop}%
\bibitem [{\citenamefont {Fernandes}\ \emph {et~al.}(2022)\citenamefont {Fernandes}, \citenamefont {Carrilho}, \citenamefont {Clifton},\ and\ \citenamefont {Mulryne}}]{Fernandes:2022zrq}%
  \BibitemOpen
  \bibfield  {author} {\bibinfo {author} {\bibfnamefont {P.~G.~S.}\ \bibnamefont {Fernandes}}, \bibinfo {author} {\bibfnamefont {P.}~\bibnamefont {Carrilho}}, \bibinfo {author} {\bibfnamefont {T.}~\bibnamefont {Clifton}},\ and\ \bibinfo {author} {\bibfnamefont {D.~J.}\ \bibnamefont {Mulryne}},\ }\bibfield  {title} {\bibinfo {title} {{The 4D Einstein-Gauss-Bonnet theory of gravity: a review}},\ }\href {https://doi.org/10.1088/1361-6382/ac500a} {\bibfield  {journal} {\bibinfo  {journal} {Class. Quant. Grav.}\ }\textbf {\bibinfo {volume} {39}},\ \bibinfo {pages} {063001} (\bibinfo {year} {2022})},\ \Eprint {https://arxiv.org/abs/2202.13908} {arXiv:2202.13908 [gr-qc]} \BibitemShut {NoStop}%
\bibitem [{\citenamefont {Horndeski}(1974)}]{Horndeski:1974wa}%
  \BibitemOpen
  \bibfield  {author} {\bibinfo {author} {\bibfnamefont {G.~W.}\ \bibnamefont {Horndeski}},\ }\bibfield  {title} {\bibinfo {title} {{Second-order scalar-tensor field equations in a four-dimensional space}},\ }\href {https://doi.org/10.1007/BF01807638} {\bibfield  {journal} {\bibinfo  {journal} {Int. J. Theor. Phys.}\ }\textbf {\bibinfo {volume} {10}},\ \bibinfo {pages} {363} (\bibinfo {year} {1974})}\BibitemShut {NoStop}%
\bibitem [{\citenamefont {Kobayashi}(2019)}]{Kobayashi:2019hrl}%
  \BibitemOpen
  \bibfield  {author} {\bibinfo {author} {\bibfnamefont {T.}~\bibnamefont {Kobayashi}},\ }\bibfield  {title} {\bibinfo {title} {{Horndeski theory and beyond: a review}},\ }\href {https://doi.org/10.1088/1361-6633/ab2429} {\bibfield  {journal} {\bibinfo  {journal} {Rept. Prog. Phys.}\ }\textbf {\bibinfo {volume} {82}},\ \bibinfo {pages} {086901} (\bibinfo {year} {2019})},\ \Eprint {https://arxiv.org/abs/1901.07183} {arXiv:1901.07183 [gr-qc]} \BibitemShut {NoStop}%
\bibitem [{\citenamefont {Mann}\ and\ \citenamefont {Ross}(1993)}]{Mann:1992ar}%
  \BibitemOpen
  \bibfield  {author} {\bibinfo {author} {\bibfnamefont {R.~B.}\ \bibnamefont {Mann}}\ and\ \bibinfo {author} {\bibfnamefont {S.~F.}\ \bibnamefont {Ross}},\ }\bibfield  {title} {\bibinfo {title} {{The D ---\ensuremath{>} 2 limit of general relativity}},\ }\href {https://doi.org/10.1088/0264-9381/10/7/015} {\bibfield  {journal} {\bibinfo  {journal} {Class. Quant. Grav.}\ }\textbf {\bibinfo {volume} {10}},\ \bibinfo {pages} {1405} (\bibinfo {year} {1993})},\ \Eprint {https://arxiv.org/abs/gr-qc/9208004} {arXiv:gr-qc/9208004} \BibitemShut {NoStop}%
\bibitem [{\citenamefont {Coll\'eaux}(2020)}]{Colleaux:2020wfv}%
  \BibitemOpen
  \bibfield  {author} {\bibinfo {author} {\bibfnamefont {A.}~\bibnamefont {Coll\'eaux}},\ }\href@noop {} {\bibinfo {title} {{Dimensional aspects of Lovelock-Lanczos gravity}}} (\bibinfo {year} {2020}),\ \Eprint {https://arxiv.org/abs/2010.14174} {arXiv:2010.14174 [gr-qc]} \BibitemShut {NoStop}%
\bibitem [{\citenamefont {Hennigar}\ \emph {et~al.}(2020{\natexlab{b}})\citenamefont {Hennigar}, \citenamefont {Kubiznak}, \citenamefont {Mann},\ and\ \citenamefont {Pollack}}]{Hennigar:2020fkv}%
  \BibitemOpen
  \bibfield  {author} {\bibinfo {author} {\bibfnamefont {R.~A.}\ \bibnamefont {Hennigar}}, \bibinfo {author} {\bibfnamefont {D.}~\bibnamefont {Kubiznak}}, \bibinfo {author} {\bibfnamefont {R.~B.}\ \bibnamefont {Mann}},\ and\ \bibinfo {author} {\bibfnamefont {C.}~\bibnamefont {Pollack}},\ }\bibfield  {title} {\bibinfo {title} {{Lower-dimensional Gauss\textendash{}Bonnet gravity and BTZ black holes}},\ }\href {https://doi.org/10.1016/j.physletb.2020.135657} {\bibfield  {journal} {\bibinfo  {journal} {Phys. Lett. B}\ }\textbf {\bibinfo {volume} {808}},\ \bibinfo {pages} {135657} (\bibinfo {year} {2020}{\natexlab{b}})},\ \Eprint {https://arxiv.org/abs/2004.12995} {arXiv:2004.12995 [gr-qc]} \BibitemShut {NoStop}%
\bibitem [{\citenamefont {Hennigar}\ \emph {et~al.}(2021)\citenamefont {Hennigar}, \citenamefont {Kubiznak},\ and\ \citenamefont {Mann}}]{Hennigar:2020drx}%
  \BibitemOpen
  \bibfield  {author} {\bibinfo {author} {\bibfnamefont {R.~A.}\ \bibnamefont {Hennigar}}, \bibinfo {author} {\bibfnamefont {D.}~\bibnamefont {Kubiznak}},\ and\ \bibinfo {author} {\bibfnamefont {R.~B.}\ \bibnamefont {Mann}},\ }\bibfield  {title} {\bibinfo {title} {{Rotating Gauss-Bonnet BTZ Black Holes}},\ }\href {https://doi.org/10.1088/1361-6382/abce48} {\bibfield  {journal} {\bibinfo  {journal} {Class. Quant. Grav.}\ }\textbf {\bibinfo {volume} {38}},\ \bibinfo {pages} {03LT01} (\bibinfo {year} {2021})},\ \Eprint {https://arxiv.org/abs/2005.13732} {arXiv:2005.13732 [gr-qc]} \BibitemShut {NoStop}%
\bibitem [{\citenamefont {Pujolas}\ \emph {et~al.}(2011)\citenamefont {Pujolas}, \citenamefont {Sawicki},\ and\ \citenamefont {Vikman}}]{Pujolas:2011he}%
  \BibitemOpen
  \bibfield  {author} {\bibinfo {author} {\bibfnamefont {O.}~\bibnamefont {Pujolas}}, \bibinfo {author} {\bibfnamefont {I.}~\bibnamefont {Sawicki}},\ and\ \bibinfo {author} {\bibfnamefont {A.}~\bibnamefont {Vikman}},\ }\bibfield  {title} {\bibinfo {title} {{The Imperfect Fluid behind Kinetic Gravity Braiding}},\ }\href {https://doi.org/10.1007/JHEP11(2011)156} {\bibfield  {journal} {\bibinfo  {journal} {JHEP}\ }\textbf {\bibinfo {volume} {11}},\ \bibinfo {pages} {156}},\ \Eprint {https://arxiv.org/abs/1103.5360} {arXiv:1103.5360 [hep-th]} \BibitemShut {NoStop}%
\bibitem [{\citenamefont {Deffayet}\ \emph {et~al.}(2010)\citenamefont {Deffayet}, \citenamefont {Pujolas}, \citenamefont {Sawicki},\ and\ \citenamefont {Vikman}}]{Deffayet:2010qz}%
  \BibitemOpen
  \bibfield  {author} {\bibinfo {author} {\bibfnamefont {C.}~\bibnamefont {Deffayet}}, \bibinfo {author} {\bibfnamefont {O.}~\bibnamefont {Pujolas}}, \bibinfo {author} {\bibfnamefont {I.}~\bibnamefont {Sawicki}},\ and\ \bibinfo {author} {\bibfnamefont {A.}~\bibnamefont {Vikman}},\ }\bibfield  {title} {\bibinfo {title} {{Imperfect Dark Energy from Kinetic Gravity Braiding}},\ }\href {https://doi.org/10.1088/1475-7516/2010/10/026} {\bibfield  {journal} {\bibinfo  {journal} {JCAP}\ }\textbf {\bibinfo {volume} {10}},\ \bibinfo {pages} {026}},\ \Eprint {https://arxiv.org/abs/1008.0048} {arXiv:1008.0048 [hep-th]} \BibitemShut {NoStop}%
\bibitem [{\citenamefont {Germani}\ and\ \citenamefont {Martin-Moruno}(2017)}]{Germani:2017pwt}%
  \BibitemOpen
  \bibfield  {author} {\bibinfo {author} {\bibfnamefont {C.}~\bibnamefont {Germani}}\ and\ \bibinfo {author} {\bibfnamefont {P.}~\bibnamefont {Martin-Moruno}},\ }\bibfield  {title} {\bibinfo {title} {{Tracking our Universe to de Sitter by a Horndeski scalar}},\ }\href {https://doi.org/10.1016/j.dark.2017.09.002} {\bibfield  {journal} {\bibinfo  {journal} {Phys. Dark Univ.}\ }\textbf {\bibinfo {volume} {18}},\ \bibinfo {pages} {1} (\bibinfo {year} {2017})},\ \Eprint {https://arxiv.org/abs/1707.03741} {arXiv:1707.03741 [gr-qc]} \BibitemShut {NoStop}%
\bibitem [{\citenamefont {Borislavov~Vasilev}\ \emph {et~al.}(2024)\citenamefont {Borislavov~Vasilev}, \citenamefont {Bouhmadi-L\'opez},\ and\ \citenamefont {Mart\'\i{}n-Moruno}}]{BorislavovVasilev:2024loq}%
  \BibitemOpen
  \bibfield  {author} {\bibinfo {author} {\bibfnamefont {T.}~\bibnamefont {Borislavov~Vasilev}}, \bibinfo {author} {\bibfnamefont {M.}~\bibnamefont {Bouhmadi-L\'opez}},\ and\ \bibinfo {author} {\bibfnamefont {P.}~\bibnamefont {Mart\'\i{}n-Moruno}},\ }\bibfield  {title} {\bibinfo {title} {{Dark energy with a shift-symmetric scalar field: Obstacles, loophole hunting and dead ends}},\ }\href {https://doi.org/10.1016/j.dark.2024.101679} {\bibfield  {journal} {\bibinfo  {journal} {Phys. Dark Univ.}\ }\textbf {\bibinfo {volume} {46}},\ \bibinfo {pages} {101679} (\bibinfo {year} {2024})},\ \Eprint {https://arxiv.org/abs/2406.12576} {arXiv:2406.12576 [gr-qc]} \BibitemShut {NoStop}%
\bibitem [{\citenamefont {Creminelli}\ \emph {et~al.}(2020)\citenamefont {Creminelli}, \citenamefont {Loayza}, \citenamefont {Serra}, \citenamefont {Trincherini},\ and\ \citenamefont {Trombetta}}]{Creminelli:2020lxn}%
  \BibitemOpen
  \bibfield  {author} {\bibinfo {author} {\bibfnamefont {P.}~\bibnamefont {Creminelli}}, \bibinfo {author} {\bibfnamefont {N.}~\bibnamefont {Loayza}}, \bibinfo {author} {\bibfnamefont {F.}~\bibnamefont {Serra}}, \bibinfo {author} {\bibfnamefont {E.}~\bibnamefont {Trincherini}},\ and\ \bibinfo {author} {\bibfnamefont {L.~G.}\ \bibnamefont {Trombetta}},\ }\bibfield  {title} {\bibinfo {title} {{Hairy Black-holes in Shift-symmetric Theories}},\ }\href {https://doi.org/10.1007/JHEP08(2020)045} {\bibfield  {journal} {\bibinfo  {journal} {JHEP}\ }\textbf {\bibinfo {volume} {08}},\ \bibinfo {pages} {045}},\ \Eprint {https://arxiv.org/abs/2004.02893} {arXiv:2004.02893 [hep-th]} \BibitemShut {NoStop}%
\bibitem [{\citenamefont {Sotiriou}\ and\ \citenamefont {Zhou}(2014{\natexlab{a}})}]{Sotiriou:2013qea}%
  \BibitemOpen
  \bibfield  {author} {\bibinfo {author} {\bibfnamefont {T.~P.}\ \bibnamefont {Sotiriou}}\ and\ \bibinfo {author} {\bibfnamefont {S.-Y.}\ \bibnamefont {Zhou}},\ }\bibfield  {title} {\bibinfo {title} {{Black hole hair in generalized scalar-tensor gravity}},\ }\href {https://doi.org/10.1103/PhysRevLett.112.251102} {\bibfield  {journal} {\bibinfo  {journal} {Phys. Rev. Lett.}\ }\textbf {\bibinfo {volume} {112}},\ \bibinfo {pages} {251102} (\bibinfo {year} {2014}{\natexlab{a}})},\ \Eprint {https://arxiv.org/abs/1312.3622} {arXiv:1312.3622 [gr-qc]} \BibitemShut {NoStop}%
\bibitem [{\citenamefont {Sotiriou}\ and\ \citenamefont {Zhou}(2014{\natexlab{b}})}]{Sotiriou:2014pfa}%
  \BibitemOpen
  \bibfield  {author} {\bibinfo {author} {\bibfnamefont {T.~P.}\ \bibnamefont {Sotiriou}}\ and\ \bibinfo {author} {\bibfnamefont {S.-Y.}\ \bibnamefont {Zhou}},\ }\bibfield  {title} {\bibinfo {title} {{Black hole hair in generalized scalar-tensor gravity: An explicit example}},\ }\href {https://doi.org/10.1103/PhysRevD.90.124063} {\bibfield  {journal} {\bibinfo  {journal} {Phys. Rev. D}\ }\textbf {\bibinfo {volume} {90}},\ \bibinfo {pages} {124063} (\bibinfo {year} {2014}{\natexlab{b}})},\ \Eprint {https://arxiv.org/abs/1408.1698} {arXiv:1408.1698 [gr-qc]} \BibitemShut {NoStop}%
\bibitem [{\citenamefont {Delgado}\ \emph {et~al.}(2020)\citenamefont {Delgado}, \citenamefont {Herdeiro},\ and\ \citenamefont {Radu}}]{Delgado:2020rev}%
  \BibitemOpen
  \bibfield  {author} {\bibinfo {author} {\bibfnamefont {J.~F.~M.}\ \bibnamefont {Delgado}}, \bibinfo {author} {\bibfnamefont {C.~A.~R.}\ \bibnamefont {Herdeiro}},\ and\ \bibinfo {author} {\bibfnamefont {E.}~\bibnamefont {Radu}},\ }\bibfield  {title} {\bibinfo {title} {{Spinning black holes in shift-symmetric Horndeski theory}},\ }\href {https://doi.org/10.1007/JHEP04(2020)180} {\bibfield  {journal} {\bibinfo  {journal} {JHEP}\ }\textbf {\bibinfo {volume} {04}},\ \bibinfo {pages} {180}},\ \Eprint {https://arxiv.org/abs/2002.05012} {arXiv:2002.05012 [gr-qc]} \BibitemShut {NoStop}%
\bibitem [{\citenamefont {Khoury}\ \emph {et~al.}(2020)\citenamefont {Khoury}, \citenamefont {Trodden},\ and\ \citenamefont {Wong}}]{Khoury:2020aya}%
  \BibitemOpen
  \bibfield  {author} {\bibinfo {author} {\bibfnamefont {J.}~\bibnamefont {Khoury}}, \bibinfo {author} {\bibfnamefont {M.}~\bibnamefont {Trodden}},\ and\ \bibinfo {author} {\bibfnamefont {S.~S.~C.}\ \bibnamefont {Wong}},\ }\bibfield  {title} {\bibinfo {title} {{Existence and instability of hairy black holes in shift-symmetric Horndeski theories}},\ }\href {https://doi.org/10.1088/1475-7516/2020/11/044} {\bibfield  {journal} {\bibinfo  {journal} {JCAP}\ }\textbf {\bibinfo {volume} {11}},\ \bibinfo {pages} {044}},\ \Eprint {https://arxiv.org/abs/2007.01320} {arXiv:2007.01320 [astro-ph.CO]} \BibitemShut {NoStop}%
\bibitem [{\citenamefont {Saravani}\ and\ \citenamefont {Sotiriou}(2019)}]{Saravani:2019xwx}%
  \BibitemOpen
  \bibfield  {author} {\bibinfo {author} {\bibfnamefont {M.}~\bibnamefont {Saravani}}\ and\ \bibinfo {author} {\bibfnamefont {T.~P.}\ \bibnamefont {Sotiriou}},\ }\bibfield  {title} {\bibinfo {title} {{Classification of shift-symmetric Horndeski theories and hairy black holes}},\ }\href {https://doi.org/10.1103/PhysRevD.99.124004} {\bibfield  {journal} {\bibinfo  {journal} {Phys. Rev. D}\ }\textbf {\bibinfo {volume} {99}},\ \bibinfo {pages} {124004} (\bibinfo {year} {2019})},\ \Eprint {https://arxiv.org/abs/1903.02055} {arXiv:1903.02055 [gr-qc]} \BibitemShut {NoStop}%
\bibitem [{\citenamefont {Babichev}\ \emph {et~al.}(2017)\citenamefont {Babichev}, \citenamefont {Charmousis},\ and\ \citenamefont {Leh\'ebel}}]{Babichev:2017guv}%
  \BibitemOpen
  \bibfield  {author} {\bibinfo {author} {\bibfnamefont {E.}~\bibnamefont {Babichev}}, \bibinfo {author} {\bibfnamefont {C.}~\bibnamefont {Charmousis}},\ and\ \bibinfo {author} {\bibfnamefont {A.}~\bibnamefont {Leh\'ebel}},\ }\bibfield  {title} {\bibinfo {title} {{Asymptotically flat black holes in Horndeski theory and beyond}},\ }\href {https://doi.org/10.1088/1475-7516/2017/04/027} {\bibfield  {journal} {\bibinfo  {journal} {JCAP}\ }\textbf {\bibinfo {volume} {04}},\ \bibinfo {pages} {027}},\ \Eprint {https://arxiv.org/abs/1702.01938} {arXiv:1702.01938 [gr-qc]} \BibitemShut {NoStop}%
\bibitem [{\citenamefont {Kobayashi}\ \emph {et~al.}(2011)\citenamefont {Kobayashi}, \citenamefont {Yamaguchi},\ and\ \citenamefont {Yokoyama}}]{Kobayashi:2011nu}%
  \BibitemOpen
  \bibfield  {author} {\bibinfo {author} {\bibfnamefont {T.}~\bibnamefont {Kobayashi}}, \bibinfo {author} {\bibfnamefont {M.}~\bibnamefont {Yamaguchi}},\ and\ \bibinfo {author} {\bibfnamefont {J.}~\bibnamefont {Yokoyama}},\ }\bibfield  {title} {\bibinfo {title} {{Generalized G-inflation: Inflation with the most general second-order field equations}},\ }\href {https://doi.org/10.1143/PTP.126.511} {\bibfield  {journal} {\bibinfo  {journal} {Prog. Theor. Phys.}\ }\textbf {\bibinfo {volume} {126}},\ \bibinfo {pages} {511} (\bibinfo {year} {2011})},\ \Eprint {https://arxiv.org/abs/1105.5723} {arXiv:1105.5723 [hep-th]} \BibitemShut {NoStop}%
\bibitem [{\citenamefont {Lecoeur}(2024)}]{Lecoeur:2024kwe}%
  \BibitemOpen
  \bibfield  {author} {\bibinfo {author} {\bibfnamefont {N.}~\bibnamefont {Lecoeur}},\ }\emph {\bibinfo {title} {{Exact black hole solutions in scalar-tensor theories}}},\ \href@noop {} {Ph.D. thesis},\ \bibinfo  {school} {U. Paris-Saclay} (\bibinfo {year} {2024}),\ \Eprint {https://arxiv.org/abs/2406.11095} {arXiv:2406.11095 [gr-qc]} \BibitemShut {NoStop}%
\bibitem [{\citenamefont {Deruelle}\ and\ \citenamefont {Farina-Busto}(1990)}]{Deruelle:1989fj}%
  \BibitemOpen
  \bibfield  {author} {\bibinfo {author} {\bibfnamefont {N.}~\bibnamefont {Deruelle}}\ and\ \bibinfo {author} {\bibfnamefont {L.}~\bibnamefont {Farina-Busto}},\ }\bibfield  {title} {\bibinfo {title} {{The Lovelock Gravitational Field Equations in Cosmology}},\ }\href {https://doi.org/10.1103/PhysRevD.41.3696} {\bibfield  {journal} {\bibinfo  {journal} {Phys. Rev. D}\ }\textbf {\bibinfo {volume} {41}},\ \bibinfo {pages} {3696} (\bibinfo {year} {1990})}\BibitemShut {NoStop}%
\bibitem [{Note1()}]{Note1}%
  \BibitemOpen
  \bibinfo {note} {$c_1=1$ is the coefficient of the Einstein-Hilbert term.}\BibitemShut {Stop}%
\bibitem [{Note2()}]{Note2}%
  \BibitemOpen
  \bibinfo {note} {From the continuity equation, we obtain $\rho = \rho _0 a^{-3(1+\omega )}$, where $\omega $ is the equation of state parameter, $p = \omega \rho $. In the early universe, radiation ($\omega =1/3$) dominates over matter ($\omega =0$).}\BibitemShut {Stop}%
\bibitem [{Note3()}]{Note3}%
  \BibitemOpen
  \bibinfo {note} {The same qualitative conclusions hold when other initial conditions, and equations of state, such as matter ($\omega =0$), are considered.}\BibitemShut {Stop}%
\bibitem [{\citenamefont {Cisterna}\ \emph {et~al.}(2024)\citenamefont {Cisterna}, \citenamefont {Grandi},\ and\ \citenamefont {Oliva}}]{Cisterna:2024ksz}%
  \BibitemOpen
  \bibfield  {author} {\bibinfo {author} {\bibfnamefont {A.}~\bibnamefont {Cisterna}}, \bibinfo {author} {\bibfnamefont {N.}~\bibnamefont {Grandi}},\ and\ \bibinfo {author} {\bibfnamefont {J.}~\bibnamefont {Oliva}},\ }\bibfield  {title} {\bibinfo {title} {{de Sitter geometric inflation from dynamical singularities}},\ }\href {https://doi.org/10.1103/PhysRevD.110.084043} {\bibfield  {journal} {\bibinfo  {journal} {Phys. Rev. D}\ }\textbf {\bibinfo {volume} {110}},\ \bibinfo {pages} {084043} (\bibinfo {year} {2024})},\ \Eprint {https://arxiv.org/abs/2406.10037} {arXiv:2406.10037 [hep-th]} \BibitemShut {NoStop}%
\bibitem [{\citenamefont {Oppenheimer}\ and\ \citenamefont {Snyder}(1939)}]{PhysRev.56.455}%
  \BibitemOpen
  \bibfield  {author} {\bibinfo {author} {\bibfnamefont {J.~R.}\ \bibnamefont {Oppenheimer}}\ and\ \bibinfo {author} {\bibfnamefont {H.}~\bibnamefont {Snyder}},\ }\bibfield  {title} {\bibinfo {title} {On continued gravitational contraction},\ }\href {https://doi.org/10.1103/PhysRev.56.455} {\bibfield  {journal} {\bibinfo  {journal} {Phys. Rev.}\ }\textbf {\bibinfo {volume} {56}},\ \bibinfo {pages} {455} (\bibinfo {year} {1939})}\BibitemShut {NoStop}%
\bibitem [{\citenamefont {Cipriani}\ \emph {et~al.}(2024)\citenamefont {Cipriani}, \citenamefont {Fazzini},\ and\ \citenamefont {Wilson-Ewing}}]{Cipriani:2024nhx}%
  \BibitemOpen
  \bibfield  {author} {\bibinfo {author} {\bibfnamefont {L.}~\bibnamefont {Cipriani}}, \bibinfo {author} {\bibfnamefont {F.}~\bibnamefont {Fazzini}},\ and\ \bibinfo {author} {\bibfnamefont {E.}~\bibnamefont {Wilson-Ewing}},\ }\bibfield  {title} {\bibinfo {title} {{Gravitational collapse in effective loop quantum gravity: Beyond marginally bound configurations}},\ }\href {https://doi.org/10.1103/PhysRevD.110.066004} {\bibfield  {journal} {\bibinfo  {journal} {Phys. Rev. D}\ }\textbf {\bibinfo {volume} {110}},\ \bibinfo {pages} {066004} (\bibinfo {year} {2024})},\ \Eprint {https://arxiv.org/abs/2404.04192} {arXiv:2404.04192 [gr-qc]} \BibitemShut {NoStop}%
\bibitem [{\citenamefont {Alkac}\ \emph {et~al.}(2022)\citenamefont {Alkac}, \citenamefont {Ozen},\ and\ \citenamefont {Suer}}]{Alkac:2022fuc}%
  \BibitemOpen
  \bibfield  {author} {\bibinfo {author} {\bibfnamefont {G.}~\bibnamefont {Alkac}}, \bibinfo {author} {\bibfnamefont {G.~D.}\ \bibnamefont {Ozen}},\ and\ \bibinfo {author} {\bibfnamefont {G.}~\bibnamefont {Suer}},\ }\bibfield  {title} {\bibinfo {title} {{Lower-dimensional limits of cubic Lovelock gravity}},\ }\href {https://doi.org/10.1016/j.nuclphysb.2022.116027} {\bibfield  {journal} {\bibinfo  {journal} {Nucl. Phys. B}\ }\textbf {\bibinfo {volume} {985}},\ \bibinfo {pages} {116027} (\bibinfo {year} {2022})},\ \Eprint {https://arxiv.org/abs/2203.01811} {arXiv:2203.01811 [gr-qc]} \BibitemShut {NoStop}%
\bibitem [{\citenamefont {Birmingham}(1999)}]{Birmingham:1998nr}%
  \BibitemOpen
  \bibfield  {author} {\bibinfo {author} {\bibfnamefont {D.}~\bibnamefont {Birmingham}},\ }\bibfield  {title} {\bibinfo {title} {{Topological black holes in Anti-de Sitter space}},\ }\href {https://doi.org/10.1088/0264-9381/16/4/009} {\bibfield  {journal} {\bibinfo  {journal} {Class. Quant. Grav.}\ }\textbf {\bibinfo {volume} {16}},\ \bibinfo {pages} {1197} (\bibinfo {year} {1999})},\ \Eprint {https://arxiv.org/abs/hep-th/9808032} {arXiv:hep-th/9808032} \BibitemShut {NoStop}%
\bibitem [{\citenamefont {Lemos}(1998)}]{Lemos:1997bd}%
  \BibitemOpen
  \bibfield  {author} {\bibinfo {author} {\bibfnamefont {J.~P.~S.}\ \bibnamefont {Lemos}},\ }\bibfield  {title} {\bibinfo {title} {{Gravitational collapse to toroidal, cylindrical and planar black holes with gravitational and other forms of radiation}},\ }\href {https://doi.org/10.1103/PhysRevD.57.4600} {\bibfield  {journal} {\bibinfo  {journal} {Phys. Rev. D}\ }\textbf {\bibinfo {volume} {57}},\ \bibinfo {pages} {4600} (\bibinfo {year} {1998})},\ \Eprint {https://arxiv.org/abs/gr-qc/9709013} {arXiv:gr-qc/9709013} \BibitemShut {NoStop}%
\bibitem [{\citenamefont {Lemos}\ and\ \citenamefont {Zanchin}(1996)}]{Lemos:1995cm}%
  \BibitemOpen
  \bibfield  {author} {\bibinfo {author} {\bibfnamefont {J.~P.~S.}\ \bibnamefont {Lemos}}\ and\ \bibinfo {author} {\bibfnamefont {V.~T.}\ \bibnamefont {Zanchin}},\ }\bibfield  {title} {\bibinfo {title} {{Rotating charged black string and three-dimensional black holes}},\ }\href {https://doi.org/10.1103/PhysRevD.54.3840} {\bibfield  {journal} {\bibinfo  {journal} {Phys. Rev. D}\ }\textbf {\bibinfo {volume} {54}},\ \bibinfo {pages} {3840} (\bibinfo {year} {1996})},\ \Eprint {https://arxiv.org/abs/hep-th/9511188} {arXiv:hep-th/9511188} \BibitemShut {NoStop}%
\bibitem [{Note4()}]{Note4}%
  \BibitemOpen
  \bibinfo {note} {For $k=0$, it is possible to have toroidal, cylindrical or planar topology, depending on the range of the coordinates defining the two-dimensional space $\protect \mathrm {d}\Omega ^2_0$ \cite {Lemos:1997bd}.}\BibitemShut {Stop}%
\bibitem [{Note5()}]{Note5}%
  \BibitemOpen
  \bibinfo {note} {To our knowledge, this is the first example of a staticity theorem holding for a class of higher-derivative theories beyond GR in four-dimensions. However, see Refs.~\cite {Bueno:2025jgc,Oliva:2010eb,Oliva:2011xu,Bueno:2024eig,Bueno:2024zsx} for specific cases in $2+1$ dimensions and higher-dimensional scenarios where staticity (and in some cases, Birkhoff) theorems have been established.}\BibitemShut {Stop}%
\bibitem [{\citenamefont {de~Paula~Netto}\ \emph {et~al.}(2024)\citenamefont {de~Paula~Netto}, \citenamefont {Giacchini}, \citenamefont {Burzill\`a},\ and\ \citenamefont {Modesto}}]{dePaulaNetto:2023cjw}%
  \BibitemOpen
  \bibfield  {author} {\bibinfo {author} {\bibfnamefont {T.}~\bibnamefont {de~Paula~Netto}}, \bibinfo {author} {\bibfnamefont {B.~L.}\ \bibnamefont {Giacchini}}, \bibinfo {author} {\bibfnamefont {N.}~\bibnamefont {Burzill\`a}},\ and\ \bibinfo {author} {\bibfnamefont {L.}~\bibnamefont {Modesto}},\ }\bibfield  {title} {\bibinfo {title} {{On effective models of regular black holes inspired by higher-derivative and nonlocal gravity}},\ }\href {https://doi.org/10.1016/j.nuclphysb.2024.116674} {\bibfield  {journal} {\bibinfo  {journal} {Nucl. Phys. B}\ }\textbf {\bibinfo {volume} {1007}},\ \bibinfo {pages} {116674} (\bibinfo {year} {2024})},\ \Eprint {https://arxiv.org/abs/2308.12251} {arXiv:2308.12251 [gr-qc]} \BibitemShut {NoStop}%
\bibitem [{\citenamefont {Lobo}\ \emph {et~al.}(2021)\citenamefont {Lobo}, \citenamefont {Rodrigues}, \citenamefont {de~Sousa~Silva}, \citenamefont {Simpson},\ and\ \citenamefont {Visser}}]{Lobo:2020ffi}%
  \BibitemOpen
  \bibfield  {author} {\bibinfo {author} {\bibfnamefont {F.~S.~N.}\ \bibnamefont {Lobo}}, \bibinfo {author} {\bibfnamefont {M.~E.}\ \bibnamefont {Rodrigues}}, \bibinfo {author} {\bibfnamefont {M.~V.}\ \bibnamefont {de~Sousa~Silva}}, \bibinfo {author} {\bibfnamefont {A.}~\bibnamefont {Simpson}},\ and\ \bibinfo {author} {\bibfnamefont {M.}~\bibnamefont {Visser}},\ }\bibfield  {title} {\bibinfo {title} {{Novel black-bounce spacetimes: wormholes, regularity, energy conditions, and causal structure}},\ }\href {https://doi.org/10.1103/PhysRevD.103.084052} {\bibfield  {journal} {\bibinfo  {journal} {Phys. Rev. D}\ }\textbf {\bibinfo {volume} {103}},\ \bibinfo {pages} {084052} (\bibinfo {year} {2021})},\ \Eprint {https://arxiv.org/abs/2009.12057} {arXiv:2009.12057 [gr-qc]} \BibitemShut {NoStop}%
\bibitem [{\citenamefont {Bolokhov}\ \emph {et~al.}(2024)\citenamefont {Bolokhov}, \citenamefont {Bronnikov},\ and\ \citenamefont {Skvortsova}}]{Bolokhov:2024sdy}%
  \BibitemOpen
  \bibfield  {author} {\bibinfo {author} {\bibfnamefont {S.~V.}\ \bibnamefont {Bolokhov}}, \bibinfo {author} {\bibfnamefont {K.~A.}\ \bibnamefont {Bronnikov}},\ and\ \bibinfo {author} {\bibfnamefont {M.~V.}\ \bibnamefont {Skvortsova}},\ }\bibfield  {title} {\bibinfo {title} {{A Regular Center Instead of a Black Bounce}},\ }\href {https://doi.org/10.1134/S0202289324700178} {\bibfield  {journal} {\bibinfo  {journal} {Grav. Cosmol.}\ }\textbf {\bibinfo {volume} {30}},\ \bibinfo {pages} {265} (\bibinfo {year} {2024})},\ \Eprint {https://arxiv.org/abs/2405.09124} {arXiv:2405.09124 [gr-qc]} \BibitemShut {NoStop}%
\bibitem [{\citenamefont {Simpson}(2023)}]{Simpson:2023apa}%
  \BibitemOpen
  \bibfield  {author} {\bibinfo {author} {\bibfnamefont {A.}~\bibnamefont {Simpson}},\ }\emph {\bibinfo {title} {{Excising Curvature Singularities from General Relativity}}},\ \href {https://doi.org/10.26686/wgtn.22338298} {\bibinfo {type} {Other thesis}} (\bibinfo {year} {2023}),\ \Eprint {https://arxiv.org/abs/2304.07383} {arXiv:2304.07383 [gr-qc]} \BibitemShut {NoStop}%
\bibitem [{\citenamefont {Zhou}\ and\ \citenamefont {Modesto}(2023{\natexlab{a}})}]{Zhou:2022yio}%
  \BibitemOpen
  \bibfield  {author} {\bibinfo {author} {\bibfnamefont {T.}~\bibnamefont {Zhou}}\ and\ \bibinfo {author} {\bibfnamefont {L.}~\bibnamefont {Modesto}},\ }\bibfield  {title} {\bibinfo {title} {{Geodesic incompleteness of some popular regular black holes}},\ }\href {https://doi.org/10.1103/PhysRevD.107.044016} {\bibfield  {journal} {\bibinfo  {journal} {Phys. Rev. D}\ }\textbf {\bibinfo {volume} {107}},\ \bibinfo {pages} {044016} (\bibinfo {year} {2023}{\natexlab{a}})},\ \Eprint {https://arxiv.org/abs/2208.02557} {arXiv:2208.02557 [gr-qc]} \BibitemShut {NoStop}%
\bibitem [{\citenamefont {Zhou}\ and\ \citenamefont {Modesto}(2023{\natexlab{b}})}]{Zhou:2023lwc}%
  \BibitemOpen
  \bibfield  {author} {\bibinfo {author} {\bibfnamefont {T.}~\bibnamefont {Zhou}}\ and\ \bibinfo {author} {\bibfnamefont {L.}~\bibnamefont {Modesto}},\ }\href@noop {} {\bibinfo {title} {{On the analytic extension of regular rotating black holes}}} (\bibinfo {year} {2023}{\natexlab{b}}),\ \Eprint {https://arxiv.org/abs/2303.11322} {arXiv:2303.11322 [gr-qc]} \BibitemShut {NoStop}%
\bibitem [{\citenamefont {Carballo-Rubio}\ \emph {et~al.}(2022)\citenamefont {Carballo-Rubio}, \citenamefont {Di~Filippo}, \citenamefont {Liberati}, \citenamefont {Pacilio},\ and\ \citenamefont {Visser}}]{Carballo-Rubio:2022kad}%
  \BibitemOpen
  \bibfield  {author} {\bibinfo {author} {\bibfnamefont {R.}~\bibnamefont {Carballo-Rubio}}, \bibinfo {author} {\bibfnamefont {F.}~\bibnamefont {Di~Filippo}}, \bibinfo {author} {\bibfnamefont {S.}~\bibnamefont {Liberati}}, \bibinfo {author} {\bibfnamefont {C.}~\bibnamefont {Pacilio}},\ and\ \bibinfo {author} {\bibfnamefont {M.}~\bibnamefont {Visser}},\ }\bibfield  {title} {\bibinfo {title} {{Regular black holes without mass inflation instability}},\ }\href {https://doi.org/10.1007/JHEP09(2022)118} {\bibfield  {journal} {\bibinfo  {journal} {JHEP}\ }\textbf {\bibinfo {volume} {09}},\ \bibinfo {pages} {118}},\ \Eprint {https://arxiv.org/abs/2205.13556} {arXiv:2205.13556 [gr-qc]} \BibitemShut {NoStop}%
\bibitem [{\citenamefont {Carballo-Rubio}\ \emph {et~al.}(2024)\citenamefont {Carballo-Rubio}, \citenamefont {Di~Filippo}, \citenamefont {Liberati},\ and\ \citenamefont {Visser}}]{Carballo-Rubio:2024dca}%
  \BibitemOpen
  \bibfield  {author} {\bibinfo {author} {\bibfnamefont {R.}~\bibnamefont {Carballo-Rubio}}, \bibinfo {author} {\bibfnamefont {F.}~\bibnamefont {Di~Filippo}}, \bibinfo {author} {\bibfnamefont {S.}~\bibnamefont {Liberati}},\ and\ \bibinfo {author} {\bibfnamefont {M.}~\bibnamefont {Visser}},\ }\bibfield  {title} {\bibinfo {title} {{Mass Inflation without Cauchy Horizons}},\ }\href {https://doi.org/10.1103/PhysRevLett.133.181402} {\bibfield  {journal} {\bibinfo  {journal} {Phys. Rev. Lett.}\ }\textbf {\bibinfo {volume} {133}},\ \bibinfo {pages} {181402} (\bibinfo {year} {2024})},\ \Eprint {https://arxiv.org/abs/2402.14913} {arXiv:2402.14913 [gr-qc]} \BibitemShut {NoStop}%
\bibitem [{\citenamefont {Franzin}\ \emph {et~al.}(2022)\citenamefont {Franzin}, \citenamefont {Liberati}, \citenamefont {Mazza},\ and\ \citenamefont {Vellucci}}]{Franzin:2022wai}%
  \BibitemOpen
  \bibfield  {author} {\bibinfo {author} {\bibfnamefont {E.}~\bibnamefont {Franzin}}, \bibinfo {author} {\bibfnamefont {S.}~\bibnamefont {Liberati}}, \bibinfo {author} {\bibfnamefont {J.}~\bibnamefont {Mazza}},\ and\ \bibinfo {author} {\bibfnamefont {V.}~\bibnamefont {Vellucci}},\ }\bibfield  {title} {\bibinfo {title} {{Stable rotating regular black holes}},\ }\href {https://doi.org/10.1103/PhysRevD.106.104060} {\bibfield  {journal} {\bibinfo  {journal} {Phys. Rev. D}\ }\textbf {\bibinfo {volume} {106}},\ \bibinfo {pages} {104060} (\bibinfo {year} {2022})},\ \Eprint {https://arxiv.org/abs/2207.08864} {arXiv:2207.08864 [gr-qc]} \BibitemShut {NoStop}%
\bibitem [{\citenamefont {Bueno}\ \emph {et~al.}(2021)\citenamefont {Bueno}, \citenamefont {Cano}, \citenamefont {Moreno},\ and\ \citenamefont {van~der Velde}}]{Bueno:2021krl}%
  \BibitemOpen
  \bibfield  {author} {\bibinfo {author} {\bibfnamefont {P.}~\bibnamefont {Bueno}}, \bibinfo {author} {\bibfnamefont {P.~A.}\ \bibnamefont {Cano}}, \bibinfo {author} {\bibfnamefont {J.}~\bibnamefont {Moreno}},\ and\ \bibinfo {author} {\bibfnamefont {G.}~\bibnamefont {van~der Velde}},\ }\bibfield  {title} {\bibinfo {title} {{Regular black holes in three dimensions}},\ }\href {https://doi.org/10.1103/PhysRevD.104.L021501} {\bibfield  {journal} {\bibinfo  {journal} {Phys. Rev. D}\ }\textbf {\bibinfo {volume} {104}},\ \bibinfo {pages} {L021501} (\bibinfo {year} {2021})},\ \Eprint {https://arxiv.org/abs/2104.10172} {arXiv:2104.10172 [gr-qc]} \BibitemShut {NoStop}%
\bibitem [{\citenamefont {Bueno}\ \emph {et~al.}(2025{\natexlab{b}})\citenamefont {Bueno}, \citenamefont {Lasso~Andino}, \citenamefont {Moreno},\ and\ \citenamefont {van~der Velde}}]{Bueno:2025jgc}%
  \BibitemOpen
  \bibfield  {author} {\bibinfo {author} {\bibfnamefont {P.}~\bibnamefont {Bueno}}, \bibinfo {author} {\bibfnamefont {O.}~\bibnamefont {Lasso~Andino}}, \bibinfo {author} {\bibfnamefont {J.}~\bibnamefont {Moreno}},\ and\ \bibinfo {author} {\bibfnamefont {G.}~\bibnamefont {van~der Velde}},\ }\href@noop {} {\bibinfo {title} {{On regular charged black holes in three dimensions}}} (\bibinfo {year} {2025}{\natexlab{b}}),\ \Eprint {https://arxiv.org/abs/2503.02930} {arXiv:2503.02930 [gr-qc]} \BibitemShut {NoStop}%
\bibitem [{\citenamefont {Husain}\ \emph {et~al.}(2022)\citenamefont {Husain}, \citenamefont {Kelly}, \citenamefont {Santacruz},\ and\ \citenamefont {Wilson-Ewing}}]{Husain:2021ojz}%
  \BibitemOpen
  \bibfield  {author} {\bibinfo {author} {\bibfnamefont {V.}~\bibnamefont {Husain}}, \bibinfo {author} {\bibfnamefont {J.~G.}\ \bibnamefont {Kelly}}, \bibinfo {author} {\bibfnamefont {R.}~\bibnamefont {Santacruz}},\ and\ \bibinfo {author} {\bibfnamefont {E.}~\bibnamefont {Wilson-Ewing}},\ }\bibfield  {title} {\bibinfo {title} {{Quantum Gravity of Dust Collapse: Shock Waves from Black Holes}},\ }\href {https://doi.org/10.1103/PhysRevLett.128.121301} {\bibfield  {journal} {\bibinfo  {journal} {Phys. Rev. Lett.}\ }\textbf {\bibinfo {volume} {128}},\ \bibinfo {pages} {121301} (\bibinfo {year} {2022})},\ \Eprint {https://arxiv.org/abs/2109.08667} {arXiv:2109.08667 [gr-qc]} \BibitemShut {NoStop}%
\bibitem [{\citenamefont {Lewandowski}\ \emph {et~al.}(2023)\citenamefont {Lewandowski}, \citenamefont {Ma}, \citenamefont {Yang},\ and\ \citenamefont {Zhang}}]{Lewandowski:2022zce}%
  \BibitemOpen
  \bibfield  {author} {\bibinfo {author} {\bibfnamefont {J.}~\bibnamefont {Lewandowski}}, \bibinfo {author} {\bibfnamefont {Y.}~\bibnamefont {Ma}}, \bibinfo {author} {\bibfnamefont {J.}~\bibnamefont {Yang}},\ and\ \bibinfo {author} {\bibfnamefont {C.}~\bibnamefont {Zhang}},\ }\bibfield  {title} {\bibinfo {title} {{Quantum Oppenheimer-Snyder and Swiss Cheese Models}},\ }\href {https://doi.org/10.1103/PhysRevLett.130.101501} {\bibfield  {journal} {\bibinfo  {journal} {Phys. Rev. Lett.}\ }\textbf {\bibinfo {volume} {130}},\ \bibinfo {pages} {101501} (\bibinfo {year} {2023})},\ \Eprint {https://arxiv.org/abs/2210.02253} {arXiv:2210.02253 [gr-qc]} \BibitemShut {NoStop}%
\bibitem [{\citenamefont {Bonanno}\ \emph {et~al.}(2024)\citenamefont {Bonanno}, \citenamefont {Malafarina},\ and\ \citenamefont {Panassiti}}]{Bonanno:2023rzk}%
  \BibitemOpen
  \bibfield  {author} {\bibinfo {author} {\bibfnamefont {A.}~\bibnamefont {Bonanno}}, \bibinfo {author} {\bibfnamefont {D.}~\bibnamefont {Malafarina}},\ and\ \bibinfo {author} {\bibfnamefont {A.}~\bibnamefont {Panassiti}},\ }\bibfield  {title} {\bibinfo {title} {{Dust Collapse in Asymptotic Safety: A Path to Regular Black Holes}},\ }\href {https://doi.org/10.1103/PhysRevLett.132.031401} {\bibfield  {journal} {\bibinfo  {journal} {Phys. Rev. Lett.}\ }\textbf {\bibinfo {volume} {132}},\ \bibinfo {pages} {031401} (\bibinfo {year} {2024})},\ \Eprint {https://arxiv.org/abs/2308.10890} {arXiv:2308.10890 [gr-qc]} \BibitemShut {NoStop}%
\bibitem [{\citenamefont {de~Cesare}\ \emph {et~al.}(2020)\citenamefont {de~Cesare}, \citenamefont {Seahra},\ and\ \citenamefont {Wilson-Ewing}}]{deCesare:2020swb}%
  \BibitemOpen
  \bibfield  {author} {\bibinfo {author} {\bibfnamefont {M.}~\bibnamefont {de~Cesare}}, \bibinfo {author} {\bibfnamefont {S.~S.}\ \bibnamefont {Seahra}},\ and\ \bibinfo {author} {\bibfnamefont {E.}~\bibnamefont {Wilson-Ewing}},\ }\bibfield  {title} {\bibinfo {title} {{The singularity in mimetic Kantowski-Sachs cosmology}},\ }\href {https://doi.org/10.1088/1475-7516/2020/07/018} {\bibfield  {journal} {\bibinfo  {journal} {JCAP}\ }\textbf {\bibinfo {volume} {07}},\ \bibinfo {pages} {018}},\ \Eprint {https://arxiv.org/abs/2002.11658} {arXiv:2002.11658 [gr-qc]} \BibitemShut {NoStop}%
\bibitem [{\citenamefont {Kobayashi}(2016)}]{Kobayashi:2016xpl}%
  \BibitemOpen
  \bibfield  {author} {\bibinfo {author} {\bibfnamefont {T.}~\bibnamefont {Kobayashi}},\ }\bibfield  {title} {\bibinfo {title} {{Generic instabilities of nonsingular cosmologies in Horndeski theory: A no-go theorem}},\ }\href {https://doi.org/10.1103/PhysRevD.94.043511} {\bibfield  {journal} {\bibinfo  {journal} {Phys. Rev. D}\ }\textbf {\bibinfo {volume} {94}},\ \bibinfo {pages} {043511} (\bibinfo {year} {2016})},\ \Eprint {https://arxiv.org/abs/1606.05831} {arXiv:1606.05831 [hep-th]} \BibitemShut {NoStop}%
\bibitem [{\citenamefont {Libanov}\ \emph {et~al.}(2016)\citenamefont {Libanov}, \citenamefont {Mironov},\ and\ \citenamefont {Rubakov}}]{Libanov:2016kfc}%
  \BibitemOpen
  \bibfield  {author} {\bibinfo {author} {\bibfnamefont {M.}~\bibnamefont {Libanov}}, \bibinfo {author} {\bibfnamefont {S.}~\bibnamefont {Mironov}},\ and\ \bibinfo {author} {\bibfnamefont {V.}~\bibnamefont {Rubakov}},\ }\bibfield  {title} {\bibinfo {title} {{Generalized Galileons: instabilities of bouncing and Genesis cosmologies and modified Genesis}},\ }\href {https://doi.org/10.1088/1475-7516/2016/08/037} {\bibfield  {journal} {\bibinfo  {journal} {JCAP}\ }\textbf {\bibinfo {volume} {08}},\ \bibinfo {pages} {037}},\ \Eprint {https://arxiv.org/abs/1605.05992} {arXiv:1605.05992 [hep-th]} \BibitemShut {NoStop}%
\bibitem [{\citenamefont {Hennigar}(2017)}]{Hennigar:2017umz}%
  \BibitemOpen
  \bibfield  {author} {\bibinfo {author} {\bibfnamefont {R.~A.}\ \bibnamefont {Hennigar}},\ }\bibfield  {title} {\bibinfo {title} {{Criticality for charged black branes}},\ }\href {https://doi.org/10.1007/JHEP09(2017)082} {\bibfield  {journal} {\bibinfo  {journal} {JHEP}\ }\textbf {\bibinfo {volume} {09}},\ \bibinfo {pages} {082}},\ \Eprint {https://arxiv.org/abs/1705.07094} {arXiv:1705.07094 [hep-th]} \BibitemShut {NoStop}%
\bibitem [{\citenamefont {Cadoni}\ \emph {et~al.}(2016)\citenamefont {Cadoni}, \citenamefont {Frassino},\ and\ \citenamefont {Tuveri}}]{Cadoni:2016hhd}%
  \BibitemOpen
  \bibfield  {author} {\bibinfo {author} {\bibfnamefont {M.}~\bibnamefont {Cadoni}}, \bibinfo {author} {\bibfnamefont {A.~M.}\ \bibnamefont {Frassino}},\ and\ \bibinfo {author} {\bibfnamefont {M.}~\bibnamefont {Tuveri}},\ }\bibfield  {title} {\bibinfo {title} {{On the universality of thermodynamics and $\eta/s$ ratio for the charged Lovelock black branes}},\ }\href {https://doi.org/10.1007/JHEP05(2016)101} {\bibfield  {journal} {\bibinfo  {journal} {JHEP}\ }\textbf {\bibinfo {volume} {05}},\ \bibinfo {pages} {101}},\ \Eprint {https://arxiv.org/abs/1602.05593} {arXiv:1602.05593 [hep-th]} \BibitemShut {NoStop}%
\bibitem [{\citenamefont {Peca}\ and\ \citenamefont {Lemos}(2000)}]{Peca:1998dv}%
  \BibitemOpen
  \bibfield  {author} {\bibinfo {author} {\bibfnamefont {C.~S.}\ \bibnamefont {Peca}}\ and\ \bibinfo {author} {\bibfnamefont {J.~P.~S.}\ \bibnamefont {Lemos}},\ }\bibfield  {title} {\bibinfo {title} {{Thermodynamics of toroidal black holes}},\ }\href {https://doi.org/10.1063/1.533378} {\bibfield  {journal} {\bibinfo  {journal} {J. Math. Phys.}\ }\textbf {\bibinfo {volume} {41}},\ \bibinfo {pages} {4783} (\bibinfo {year} {2000})},\ \Eprint {https://arxiv.org/abs/gr-qc/9809029} {arXiv:gr-qc/9809029} \BibitemShut {NoStop}%
\bibitem [{\citenamefont {Hartnoll}(2009)}]{Hartnoll:2009sz}%
  \BibitemOpen
  \bibfield  {author} {\bibinfo {author} {\bibfnamefont {S.~A.}\ \bibnamefont {Hartnoll}},\ }\bibfield  {title} {\bibinfo {title} {{Lectures on holographic methods for condensed matter physics}},\ }\href {https://doi.org/10.1088/0264-9381/26/22/224002} {\bibfield  {journal} {\bibinfo  {journal} {Class. Quant. Grav.}\ }\textbf {\bibinfo {volume} {26}},\ \bibinfo {pages} {224002} (\bibinfo {year} {2009})},\ \Eprint {https://arxiv.org/abs/0903.3246} {arXiv:0903.3246 [hep-th]} \BibitemShut {NoStop}%
\bibitem [{\citenamefont {Hossenfelder}(2015)}]{Hossenfelder:2014gwa}%
  \BibitemOpen
  \bibfield  {author} {\bibinfo {author} {\bibfnamefont {S.}~\bibnamefont {Hossenfelder}},\ }\bibfield  {title} {\bibinfo {title} {{Analog Systems for Gravity Duals}},\ }\href {https://doi.org/10.1103/PhysRevD.91.124064} {\bibfield  {journal} {\bibinfo  {journal} {Phys. Rev. D}\ }\textbf {\bibinfo {volume} {91}},\ \bibinfo {pages} {124064} (\bibinfo {year} {2015})},\ \Eprint {https://arxiv.org/abs/1412.4220} {arXiv:1412.4220 [gr-qc]} \BibitemShut {NoStop}%
\bibitem [{\citenamefont {Oliva}\ and\ \citenamefont {Ray}(2010)}]{Oliva:2010eb}%
  \BibitemOpen
  \bibfield  {author} {\bibinfo {author} {\bibfnamefont {J.}~\bibnamefont {Oliva}}\ and\ \bibinfo {author} {\bibfnamefont {S.}~\bibnamefont {Ray}},\ }\bibfield  {title} {\bibinfo {title} {{A new cubic theory of gravity in five dimensions: Black hole, Birkhoff's theorem and C-function}},\ }\href {https://doi.org/10.1088/0264-9381/27/22/225002} {\bibfield  {journal} {\bibinfo  {journal} {Class. Quant. Grav.}\ }\textbf {\bibinfo {volume} {27}},\ \bibinfo {pages} {225002} (\bibinfo {year} {2010})},\ \Eprint {https://arxiv.org/abs/1003.4773} {arXiv:1003.4773 [gr-qc]} \BibitemShut {NoStop}%
\bibitem [{\citenamefont {Oliva}\ and\ \citenamefont {Ray}(2011)}]{Oliva:2011xu}%
  \BibitemOpen
  \bibfield  {author} {\bibinfo {author} {\bibfnamefont {J.}~\bibnamefont {Oliva}}\ and\ \bibinfo {author} {\bibfnamefont {S.}~\bibnamefont {Ray}},\ }\bibfield  {title} {\bibinfo {title} {{Birkhoff's Theorem in Higher Derivative Theories of Gravity}},\ }\href {https://doi.org/10.1088/0264-9381/28/17/175007} {\bibfield  {journal} {\bibinfo  {journal} {Class. Quant. Grav.}\ }\textbf {\bibinfo {volume} {28}},\ \bibinfo {pages} {175007} (\bibinfo {year} {2011})},\ \Eprint {https://arxiv.org/abs/1104.1205} {arXiv:1104.1205 [gr-qc]} \BibitemShut {NoStop}%
\end{thebibliography}%

\onecolumngrid
\newpage
\appendix

\renewcommand{\appendixname}{Supplemental Material}
\section*{SUPPLEMENTAL MATERIAL}
\setcounter{equation}{0}
\renewcommand{\theequation}{SM.\arabic{equation}}

\section{Other examples of theories}
\begin{table}[h]
    \centering
\[
\begin{array}{|c|c|c|c|c|}
    \hline
    c_n & \text{Horndeski Functions} & F(x) & H^2 & f(r) \\
    \hline
    \hline
    \frac{1-(-1)^n}{2n} &
    \begin{aligned}
        &G_2 = -2\Lambda + \frac{4X \left(28\ell^4 X^2 - 3\right)}{(1-4 \ell^4 X^2)^2} + \frac{6\tanh^{-1}\left(2\ell^2 X\right)}{\ell^2} \\&
        G_3 = \frac{2\left( 1 - 20\ell^4 X^2 \right)}{\left(1-4\ell^4 X^2\right)^2}\\&
        G_4 = \left(1-4\ell^4 X^2\right)^{-1}\\&
        G_5 = -4\ell^2 \left[ \frac{\ell^2 X}{1-4\ell^4 X^2} + \frac{1}{2} \tanh^{-1} \left( 2\ell^2 X \right) \right]
    \end{aligned}
    & \frac{\tanh^{-1}\left(\ell^2 x\right)}{\ell^2} & \frac{\tanh\left[ 8\pi \ell^2\rho/3 \right]}{\ell^2} & \frac{r^2}{\ell^2} \tanh \left[ \frac{\ell^2}{r^2} f_{\rm GR} \right]\\
    %
    %
    \hline
    1 &
    \begin{aligned}
        &G_2 = -2\Lambda + \frac{8\ell^2 X^2 \left(1+6 \ell^2 X\right)}{\left(1-2\ell^2 X\right)^3}\\&
        G_3 = \frac{2\left( 1 - 10\ell^2 X \right)}{\left(1-2\ell^2 X\right)^3}\\&
        G_4 = \left(1-2\ell^2 X\right)^{-2}\\&
        G_5 = -4\ell^2 \left( \frac{1-2\ell^4 X^2}{\left(1-2\ell^2 X\right)^2} + \log \left( \frac{2\ell^2 X}{1-2\ell^2 X} \right) \right)
    \end{aligned}
    & \frac{x}{1-\ell^2 x} & \frac{8\pi \rho}{3+\ell^2 8\pi \rho} & \frac{r^2 f_{\rm GR}}{r^2 - \ell^2 f_{\rm GR}}\\
    \hline
    \begin{aligned}&\frac{\Gamma(n/m) \delta_{0,k}}{\Gamma(1/m)\Gamma((n+m-1)/m)},\\& k = (n-1)\bmod m\end{aligned} &
    \begin{aligned}
        &G_2 = -2\Lambda + \frac{2^{m+2}\ell^{2m}X^{m+1}\left( 2m-1+3 \ell^{2m}X^m 2^m \right)}{(1-2^m \ell^{2m}X^m)^{(2m+1)/m}} \\&
        G_3 = 2\frac{1-2^m (3+2m) \ell^{2m} X^m}{(1-2^m\ell^{2m} X^m)^{(2m+1)/m}}\\&
        G_4 = \left(1-2^m\ell^{2m} X^m\right)^{-(m+1)/m}\\&
        G_5 = - \int \frac{2^m (m+1) \ell^{2m}X^{m-2}}{(1-2^m \ell^{2m} X^m)^{(2m+1)/m}} \mathrm{d}X
    \end{aligned}
    & \frac{x}{\left(1-\ell^{2m} x^m\right)^{1/m}} & \frac{8\pi \rho/3}{\left( 1 + \left(\frac{8\pi \rho}{3}\ell^2\right)^m \right)^\frac{1}{m}} & \frac{f_{\rm GR}}{\left( 1 + \left[-\frac{\ell^2}{r^2} f_{\rm GR}\right]^m \right)^{\frac{1}{m}}}\\
    \hline
    1/n &
    \begin{aligned}
        &G_2 = -2\Lambda + \frac{4 X \left(10 \ell^2 X-3\right)}{\left(1-2 \ell^2 X\right)^2}-\frac{6 \log \left(1-2 \ell^2 X\right)}{\ell^2} \\&
        G_3 = \frac{2 \left(1-6 \ell^2 X\right)}{\left(1-2 \ell^2 X\right)^2}\\&
        G_4 = (1-2\ell^2 X)^{-1}\\&
        G_5 = -2 \ell^2 \left(\frac{1}{1-2 \ell^2 X} - 2 \tanh ^{-1}\left(1-4 \ell^2 X\right)\right)
    \end{aligned}
    & -\frac{\log \left(1-\ell^2 x\right)}{\ell^2} & \frac{\left( 1 - e^{-8\pi \rho \ell^2/3} \right)}{\ell^2} & \frac{r^2}{\ell^2} \left( e^{\frac{\ell^2}{r^2}f_{\rm GR}} - 1\right)\\
    \hline
\end{array}
\]
\caption{Examples of re-summed Horndeski theories, together with their generalized Friedmann equations and the metric function for a planar black hole. We have absorbed the cosmological constant $\Lambda$ in the total energy density $\rho$. The example in the second row is a particular case ($m=1$) of the example presented in the third row.}
\end{table}

\section{Scalar field current for black holes with $k\neq 0$, and Carter-Penrose diagram for a regular black hole}
\label{app:bhs}

We have been unable to integrate the field equations and get exact solutions when $k\neq 0$ in the line-element \eqref{eq:SSS}. In the simplest case, where we consider no time dependence in the metric functions from the onset, examining the only non-trivial component of the shift-symmetry current we find for each $n$
\begin{equation}
    J^{r(n)} = \frac{2 c_n (n-1) n \ell^{2 n-2} \left(-f \phi'^2\right)^{n-2} \left(2 f N' + N(f'-2 f \phi')\right) \left(k-(2 n-3) f \left(1-r \phi'\right)^2\right)}{r^2 N}.
\end{equation}
When $k=0$, it is clear that the scalar field profile in Eq. \eqref{eq:scalarSSS} imposes $J^\mu=0$ at all orders $n$, from which the construction of planar black holes follows. Note that the other possible profile, $\phi=\frac{1}{2}\log N^2 f$, is not regular on the would-be horizon, as can be observed by examining the kinetic term $X$.
Therefore, when $k\neq 0$ the only regular scalar field profile becomes $n$ dependent. Thus, when an infinite tower is considered we obtain a highly non-linear equation for $\phi$, which when substituted in the condition to determine $f(r)$ results in an equation we have not been able to integrate. For example, for the theory with $c_n = (1-(-1)^n)/(2n)$, assuming a non-trivial and regular scalar field, we get
\begin{equation}
    k \left(\ell^4  \phi'^4 f^2-1\right)+f \left(r \phi'-1\right)^2 \left(5 \ell^4 f^2 \phi'^4+3\right) = 0,
\end{equation}
which is a sixth-order polynomial in $\phi'$. As expected, in setting $k=0$ we see that the profile \eqref{eq:scalarSSS} solves the field equation.

In Fig. \ref{fig:penrose-diagram} we present the Carter-Penrose diagram of the black hole solution \eqref{eq:BHsol1}. This diagram is generic for the geodesically complete regular black holes obtained in this framework.

\begin{figure}[]
    \centering
        \begin{tikzpicture}
        \begin{scope}[scale=0.75 ]
            \draw (0,2) -- (2,4) -- (2,0) -- cycle;
            \draw (0,2) -- (-2,4) -- (-2,0) -- cycle;

            \draw (-2,4) -- (0,6) -- (2,4);

            \coordinate (A1) at (0,2);
            \coordinate (B1) at (-2,0);
            \coordinate (C1) at (0,-2);
            \coordinate (D1) at (2,0);
            \draw (A1) -- (B1) -- (C1) -- (D1) -- cycle;

            \coordinate (L1) at (-2,-4); 
            \draw (B1) -- (L1) -- (C1) -- cycle;

            \coordinate (R1) at (2,-4); 
            \draw (D1) -- (R1) -- (C1) -- cycle;

            \coordinate (B2) at (-2,-4);
            \coordinate (C2) at (0,-6);
            \coordinate (D2) at (2,-4);
            \draw (C1) -- (B2) -- (C2) -- (D2) -- cycle;

            \coordinate (L2) at (-2,-8); 
            \draw (B2) -- (L2) -- (C2) -- cycle;

            \coordinate (R2) at (2,-8); 
            \draw (D2) -- (R2) -- (C2) -- cycle;

            \coordinate (Z1) at (-2,2);
            \coordinate (Z2) at (2,2);
            \draw (Z1) -- (B1);
            \draw (Z2) -- (D1);
            \coordinate (Z3) at (-2,-10);
            \coordinate (Z4) at (2,-10);
            \draw (-2,-8) -- (-2,-10);
            \draw (2,-8) -- (2,-10);
            
            \draw (-2,4) -- (-2,6);
            \draw (2,4) -- (2,6);


            \node at (0,0) {II};
            \node at (0,-4) {II};
            \node at (0,-8) {II};
            \node at (0,4) {II};
            \node at (1,-2) {III};
            \node at (-1,-2) {III};
            \node at (1,-6) {I};
            \node at (-1,-6) {I};
            \node at (1,2) {I};
            \node at (-1,2) {I};

            \draw (L1) -- node[midway, sloped, above] {$r=-\infty$} (B1);
            \draw (L2) -- node[midway, sloped, above] {$r=+\infty$} (B2);
            \draw (R1) -- node[midway, sloped, below] {$r=-\infty$} (D1);
            \draw (R2) -- node[midway, sloped, below] {$r=+\infty$} (D2);

            \draw (2,0) -- node[midway, sloped, below] {$r=+\infty$} (2,4);
            \draw (-2,0) -- node[midway, sloped, above] {$r=+\infty$} (-2,4);

            \draw (0,-6) -- node[midway, sloped, above] {$r_+$} (2,-4);
            \draw (0,-6) -- node[midway, sloped, above] {$r_+$} (-2,-4);

            \draw (0,-2) -- node[midway, sloped, above] {$r = 0$} (2,-4);
            \draw (0,-2) -- node[midway, sloped, above] {$r = 0$} (-2,-4);

            \draw (0,-2) -- node[midway, sloped, above] {$r = 0$} (2,0);
            \draw (0,-2) -- node[midway, sloped, above] {$r = 0$} (-2,0);
            
            \draw (0,2) -- node[midway, sloped, above] {$r_+$} (2,0);
            \draw (0,2) -- node[midway, sloped, above] {$r_+$} (-2,0);

            \draw (0,-6) -- node[midway, sloped, above] {$r_+$} (2,-8);
            \draw (0,-6) -- node[midway, sloped, above] {$r_+$} (-2,-8);

            \draw (0,-10) -- node[midway, sloped, above] {$r = 0$} (2,-8);
            \draw (0,-10) -- node[midway, sloped, above] {$r = 0$} (-2,-8);

            \draw (0,2) -- node[midway, sloped, above] {$r_+$} (2,4);
            \draw (0,2) -- node[midway, sloped, above] {$r_+$} (-2,4);

            \draw (2,4) -- node[midway, sloped, above] {$r=0$} (0,6);
            \draw (-2,4) -- node[midway, sloped, above] {$r=0$} (0,6);

        \end{scope}
        \end{tikzpicture}
    \caption{Carter-Penrose diagram of the black hole solution in Eq. \eqref{eq:BHsol1}. Region $\rm I$ corresponds to an asymptotically AdS region; Region $\rm II$ corresponds to the interior of the black hole; Region $\rm III$ corresponds to an asymptotically AdS region in a parallel universe.}
    \label{fig:penrose-diagram}
\end{figure}

\section{Generalized Friedmann equations for a spatially-curved FLRW universe}
In the main text we focused on the spatially-flat FLRW case. Here, we generalize our results for the spatially-curved FLRW case, with curvature $k\in \{-1,0,1\}$, whose metric is
\begin{equation}
    \D s^2 = -\D t^2 + a(t)^2 \left(\frac{\D r^2}{1- k r^2} + r^2 \D \theta^2 + r^2 \sin^2 \theta \D \varphi^2\right).
\end{equation}
Using the same scalar field profile as in the spatially-flat case, Eq. \eqref{eq:scalarflrw}, we find that the equation of motion for the scalar is identically solved, and we obtain the generalized Friedmann equations
\begin{equation}
    \frac{1}{\ell^2} \sum_{n=1}^{\infty} c_n \left(\ell^2 H^2\right) \left[ 1 + n \frac{k}{a^2 H^2} \right] = \frac{8\pi}{3}\rho.
\end{equation}
As an example, for the case studied in the main text, $c_n = (1-(-1)^n)/(2n)$, we obtain
\begin{equation}
    \frac{\tanh^{-1}\left(\ell^2 H^2\right)}{\ell^2} + \frac{k}{\left(1 - \ell^4 H^4\right)a^2} = \frac{8\pi}{3}\rho,
\end{equation}
and for $c_n=1$ we get
\begin{equation}
    \frac{H^2}{1-\ell^2 H^2} + \frac{k}{(1-\ell^2 H^2)^2 a^2} = \frac{8\pi}{3}\rho.
\end{equation}
Taking the continuity equation into account, from which we find $\rho \sim a^{-3(1+\omega)}$, or equivalently , $a \sim \rho^{-\frac{1}{3(1+\omega)}}$, it is possible to show that even in the spatially-curved FLRW case, these equations limit $H^2<1/\ell^2$, and therefore the curvature is bounded.

\end{document}